\documentclass[aps,prd,10pt,twocolumn,superscriptaddress,floatfix,nofootinbib,amsmath,amssymb,altaffilletter,preprintnumbers,longbibliography,eprint]{revtex4-2}
\pdfoutput=1

\usepackage{graphicx}
\usepackage{dcolumn}
\usepackage{upgreek}

\usepackage{bm}
\usepackage{braket}
\usepackage{dsfont, fontawesome}
\usepackage[colorlinks=true, pdfstartview=FitV,  breaklinks=true]{hyperref}
\usepackage{color}
\usepackage[dvipsnames,table]{xcolor}
\hypersetup{urlcolor=BlueViolet, citecolor=Plum, linkcolor=PineGreen}
\usepackage{cleveref}
\usepackage{comment}
\usepackage{booktabs}
\usepackage{siunitx}
\usepackage{soul}
\usepackage{bm}
\usepackage{esint}
\usepackage{lettrine}
\usepackage{Zallman}
\usepackage{float}
\usepackage{gensymb}
\usepackage{xspace}
\usepackage{textcomp}

\newcommand{\mDMeq}{\ensuremath{m_{\chi}}}

\begin{document}
\newcommand{\DC}[1]{{\bf \color{purple}{[DC: #1]}}}
\newcommand{\CM}[1]{{\bf \color{red}{[CM: #1]}}}

\preprint{KCL-PH-TH/2025-04}

\title{
Listening for ultra-heavy dark matter with underwater acoustic detectors
}

\author{Damon Cleaver}
    \email{damon.cleaver@kcl.ac.uk}
\author{Christopher McCabe}
\affiliation{Theoretical Particle Physics and Cosmology Group, Department of Physics, King's College London, Strand, London, WC2R 2LS, UK}

\author{Ciaran A.~J.~O'Hare}
\affiliation{ARC Centre of Excellence for Dark Matter Particle Physics, The University of Sydney, School of Physics, NSW 2006, Australia}


\begin{abstract}
Ultra-heavy dark matter candidates evade traditional direct detection experiments due to their low particle flux. 
We explore the potential of large underwater acoustic arrays, originally developed for ultra-high energy neutrino detection, to detect ultra-heavy dark matter interactions. 
These particles deposit energy via nuclear scattering while traversing seawater, generating thermo-acoustic waves detectable by hydrophones.
We present the first robust first-principles calculation of dark matter-induced acoustic waves, establishing a theoretical framework for signal modelling and sensitivity estimates. 
Our framework incorporates frequency-dependent attenuation effects, including viscous and chemical relaxation, not considered in previous calculations.
A sensitivity analysis for a hypothetical $100\,\mathrm{km}^3$ hydrophone array in the Mediterranean Sea demonstrates that such an array could extend sensitivity to the previously unexplored mass range of $0.1$~--~$10\,\upmu\mathrm{g}$ ($\sim10^{20}$~--~$10^{23}\,\mathrm{GeV}$), with sensitivity to both spin-independent and spin-dependent interactions. 
Our results establish acoustic detection as a complementary dark matter search method, enabling searches in existing hydrophone data and informing future detector designs.
\end{abstract}

\maketitle

\section{Introduction}

Despite extensive evidence for the existence of dark matter across various scales 
-- from galaxies~\cite{rubinRotationAndromedaNebula1970, ioccoEvidenceDarkMatter2015} and galaxy clusters~\cite{cloweDirectEmpiricalProof2006, masseyDarkMatterGravitational2010, harveyNongravitationalInteractionsDark2015} to large-scale structure~\cite{nelsonIllustrisTNGSimulationsPublic2021,DES:2021wwk} and cosmology~\cite{planckcollaborationPlanck2018Results2020} 
-- its fundamental nature remains unknown, 
even as the sensitivity of experiments searching for its non-gravitational interactions with Standard Model particles continues to increase.

The two leading classes of dark matter candidates are weak-scale, thermally produced particle dark matter models such as WIMPs~\cite{Jungman:1995df, Bertone:2004pz}, 
and wave-like dark matter candidates such as the QCD axion~\cite{pecceiMathrmCPConservation1977,weinbergNewLightBoson1978,wilczekProblemStrongInvariance1978}. 
In both scenarios, dark matter-induced events are rare due to suppressed couplings to Standard Model particles, despite the relatively large expected dark matter flux. This explains the absence of a signal in direct detection experiments or axion haloscopes.

However, there is another class of dark matter candidates requiring rare-event searches for a different reason:
not due to suppressed interactions, 
but because its mass, $m_{\chi}$, is `ultra-heavy', even approaching macroscopic values.
Since astronomical observations constrain the
dark matter \textit{energy} density rather than the \textit{number} density (see, e.g., ref.~\cite{Read:2014qva}), 
for a fixed energy density, heavier dark matter masses correspond to fewer particles and hence a lower terrestrial dark matter flux.
For example, for dark matter with a mass around the Planck scale ($M_{\rm{Pl}}\approx22 \,\upmu \mathrm{g}\approx1.2\times 10^{19}\,\mathrm{GeV}$), the flux on Earth is approximately $0.3 \left(22\,\upmu \mathrm{g}/m_{\chi}\right) \mathrm{m}^{-2}\, \mathrm{yr}^{-1}$. 
Therefore, direct detection experiments with typical \(\mathcal{O}(\mathrm{m}^2)\) dimensions are too small to probe dark matter masses beyond the 
Planck scale~\cite{bernabeiExtendedLimitsNeutral1999, adhikariFirstDirectDetection2022,XENON:2023iku,LZ:2024psa}, 
since the probability of a single particle passing through the detector is small, even if the scattering cross-section with Standard Model particles is larger than $\sim 10^{-24} \, \mathrm{cm}^2$, typical of electromagnetic or nuclear forces.

Theories predicting such high masses and large cross-sections generally involve composite dark matter candidates~\cite{Digman:2019wdm}, 
indicating the presence of a more complex dark sector.
Several models realise this scenario. These include
 nuclear dark matter~\cite{Hardy:2014mqa,Hardy:2015boa,Butcher:2016hic,Wise:2014ola,Krnjaic:2014xza} and 
the related scenario of dark nuggets~\cite{greshamNuclearStructureBound2017,greshamAstrophysicalSignaturesAsymmetric2018,greshamEarlyUniverseSynthesis2018},
nuclearites~\cite{derujulaNuclearitesNovelForm1984}, strangelets~\cite{lynnStrangeBaryonMatter1990}, and Q-balls~\cite{friemanPrimordialOriginNontopological1988, kusenkoSupersymmetricQballsDark1998}. 
Broader frameworks like dark blobs~\cite{grabowskaDetectingDarkBlobs2018} or loosely bound dark matter~\cite{Acevedo:2024lyr} have also been proposed. For a review of these and related models, see ref.~\cite{Carney:2022gse}. 

Although these models, which we will collectively call ultra-heavy dark matter models for simplicity, differ in structure and composition, they share a key feature: 
for dark matter heavier than $\mathcal{O}(10 \, \upmu \mathrm{g})$, the terrestrial flux becomes too low for conventional direct detection techniques to be effective, even when the interaction strength is significant.
Despite this challenge, several attempts have been made to constrain ultra-heavy dark matter candidates using alternative methods; as a result, a large part of the mass--cross-section parameter space is now excluded. 
Examples include constraints from the absence of lattice defects in ancient mica~\cite{PhysRevLett.52.1265, jacobsMacroDarkMatter2015, Acevedo:2021tbl}, 
radar meteor detectors~\cite{dhakalNewConstraintsmacroscopic2023}, the absence of fireballs in the atmosphere~\cite{sidhumacroscopicDarkMatter2019}, the heating of astrophysical gas clouds~\cite{bhoonahDetectingCompositeDark2021}, white dwarf thermonuclear runaways~\cite{grahamWhiteDwarfsDark2018, SinghSidhu:2019tbr}, and cosmological constraints~\cite{Dvorkin:2013cea, Nadler:2019zrb}. 
A more ominous constraint can be derived by assuming no dark matter collisions have resulted in human deaths or serious injuries~\cite{sidhuDeathSeriousInjury2020,Niedermann:2024hoa}. 
Additionally, proposed detection techniques include the search for long, straight damage tracks in geological quartz~\cite{ebadiUltraHeavyDarkMatter2021}.

 The challenge of detecting a small flux of cosmic particles is similarly encountered in the search for ultra-high energy (UHE) neutrinos. 
Detecting UHE neutrinos, with energies $\gtrsim 10^{18} \ \text{eV}$, would enhance our understanding of the origin of the highest energy particles in the Universe (see, e.g., refs.~\cite{Anchordoqui:2018qom, Vitagliano:2019yzm}). 
UHE neutrinos have a small flux requiring detectors with target volumes of tens of cubic kilometres~\cite{Engel:2001hd}. Even large-scale detectors like IceCube~\cite{icecubecollaborationIceCubeNeutrinoObservatory2017} and KM3NeT~\cite{katzKM3NeTProject2011} are too small to observe the expected flux of UHE neutrinos.
The need for dense instrumentation to detect Cherenkov light from neutrino interactions makes scaling these detectors to the required volumes impractical.

In response to these challenges, efforts have been made to develop \textit{acoustic} detection methods for UHE neutrinos using large arrays of hydrophones~\cite{learnedAcousticRadiationCharged1979, askariyanAcousticDetectionHigh1979, lahmannAcousticDetectionNeutrinos2016}.\footnote{Radio detectors designed to detect the Askaryan effect~\cite{askaryan1961excess,askaryan1965excess} provide another direction for the detection of UHE neutrinos.} 
Unlike electromagnetic radiation, sound waves experience much less attenuation in water, allowing detection with less dense instrumentation. 
Initial projects, such as the AMADEUS detector~\cite{aguilarAMADEUSAcousticNeutrino2011}, were designed as feasibility studies for larger detectors that could potentially be deployed in the Mediterranean Sea.

\begin{figure}[t!]
    \centering
    \includegraphics[width=0.95 \columnwidth]{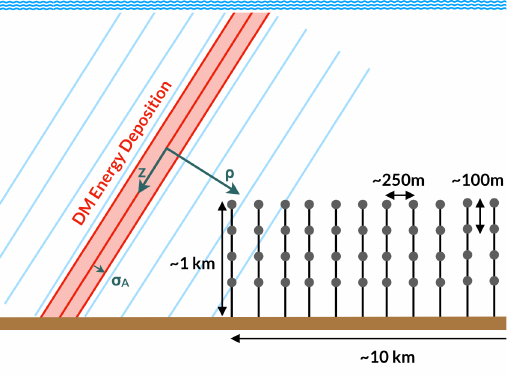}
    \caption{Acoustic detection of thermo-acoustic waves generated when ultra-heavy dark matter scatters with oxygen and hydrogen nuclei in water. The red region indicates the primary energy deposition from dark matter, which create acoustic waves that propagate outwards as cylindrical waves to an array of hydrophones (underwater pressure wave detectors) spanning a volume on the order of $100 \,\text{km}^3$. 
    Here, $\rho$ is the radial distance from the dark matter track to a hydrophone, $z$ is aligned along the centre of the energy deposition}, and~$\sigma_A$ is the width of the energy deposition profile.
    
    \label{fig:idea}
\end{figure}

Building on these acoustic detection techniques that enable the instrumentation of a large detection volume, we explore the use of an underwater array of hydrophones with a volume on the order of $100\,\mathrm{km}^3$ as detectors for ultra-heavy dark matter.
A schematic of the concept is shown in fig.~\ref{fig:idea}, which illustrates how ultra-heavy dark matter is expected to travel through water in a nearly straight line, 
radiating acoustic waves that can be detected by hydrophones. 
Building on the order-of-magnitude estimates made in ref.~\cite{grabowskaDetectingDarkBlobs2018}, 
we provide a more detailed analysis that characterises the expected signal properties, 
including the effects of attenuation, and offers refined sensitivity projections.
Our findings suggest that such an array could probe regions of the ultra-heavy dark matter parameter space that are currently unconstrained,
providing additional scientific motivation for this large-scale underwater detection method.

The paper is structured as follows: In sec.~\ref{sec:Acoustic_formalism} we derive the thermo-acoustic signal induced by dark matter and incorporate the impact of attenuation in sec.~\ref{sec:Attenuated}. We then characterise the properties of the signal post-attenuation and present sensitivity projections in sec.~\ref{sec:projections}. Finally, we conclude in sec.~\ref{sec:conclusion}. Several appendices provide technical derivations of results used in the main body of the paper.

\section{Dark Matter Induced Thermo-acoustic Radiation}\label{sec:Acoustic_formalism}

The thermo-acoustic model predicts that the near-instantaneous local heating of a liquid results in rapid expansion, which generates a pressure pulse~\cite{askariyan1957,learnedAcousticRadiationCharged1979,askariyanAcousticDetectionHigh1979}.
The thermo-acoustic model agrees well with experimental data~\cite{Lahmann:2015cdb},
and has found broad application in studies informing the design of future large-scale acoustic neutrino detectors, which, as discussed in the introduction, aim to detect the acoustic pulse produced by high-energy particle showers resulting from interactions of UHE neutrinos (see e.g., ref.~\cite{Lahmann:2019unc}).
In this section, we extend prior work in the context of UHE neutrinos, particularly that of refs.~\cite{learnedAcousticRadiationCharged1979, niessUnderwaterAcousticDetection2006}, to derive the pressure pulse produced by ultra-heavy dark matter scattering as it traverses through seawater.

We begin with the equation describing the propagation of acoustic pressure, $p(\bm{r}, t)$. In a fluid with thermal expansion coefficient $\alpha$, and specific heat capacity at constant pressure~$c_p$, $p(\bm{r}, t)$ satisfies
 \begin{equation} \label{eq:generalwave}
     \nabla^2 p - \frac{1}{c_s^2} \frac{\partial^2 p}{\partial t^2} = -\frac{\alpha}{c_p} \frac{\partial^2 q(\bm{r}, t)}{\partial t^2} \, .
 \end{equation}
where $q(\bm{r},t)$ is the energy deposition density and~$c_s$ is the speed of sound in the medium.\footnote{For non-isotropic energy depositions, an additional term that depends on the momentum transfer to the medium should be included on the right-hand-side of eq.~\eqref{eq:generalwave} (see ref.~\cite{Lahmann:2015cdb, verweij2014simulation}). We assume isotropic nuclear recoils, which results in zero net momentum transfer, so this additional term does not contribute.}
In this work, we assume parameters typical of those found in the Mediterranean Sea at a depth of 1.2\,km, a location with favourable conditions for acoustic detection experiments~\cite{Simeone:2017qwc}.
Specifically, we assume $c_s=1.52\,\mathrm{km}\,\mathrm{s}^{-1}$, $c_p=3.95\times 10^3$ $\mathrm{J \ kg^{-1} \ K^{-1}}$ and $\alpha=2.22 \times10^{-4}$ $\text{K}^{-1}$~\cite{mcdougall2011getting}, corresponding to a Gr\"uneisen parameter~$\gamma_{\mathrm{G}} = \alpha\, c_s^2/c_p= 0.13$.

The general solution to eq.~\eqref{eq:generalwave} is given by the Kirchhoff integral,
\begin{equation}
    p(\bm{r},t) = \frac{\alpha}{4\pi c_p} \int \frac{\mathrm{d}^3 \bm{r'}}{|\bm{r} - \bm{r'}|} \frac{\partial^2 q\left(\bm{r'}, t'\right)}{\partial t^2} ,
    \label{pressure_equation_full}
\end{equation}
where $t'  = t - |\bm{r} - \bm{r'}|/c_s$ is the retarded time. To make further progress, we need to determine $q(\bm{r},t)$
for ultra-heavy dark matter scattering in seawater.

\subsection{Energy deposition density}

First, we consider the time dependence of~$q(\bm{r},t)$. Given that the characteristic speed of dark matter is around $v_{\chi}\sim300\,\mathrm{km}\,\mathrm{s}^{-1}$~\cite{Lewin:1995rx}, which is significantly larger than~$c_s$,
and that the energy deposition occurs on timescales much shorter than hydrodynamic processes~\cite{Bevan:2009kd},
we assume that the energy deposition occurs instantaneously at time~$t'$. As a result, we parametrise $q(\bm{r'},t')=q(\bm{r'}) \Theta \left(t'\right)$, where $\Theta$ is the Heaviside function.\footnote{In principle, modelling the fast yet finite time for the pulse to develop is straightforward (see app.~H of ref.~\cite{niess:tel-00132273}), but as the effect is negligible, we do not include it in our analysis.
}


Next, we address the spatial dependence, which is determined by the path and energy loss of dark matter through seawater.
In this work, the typical cross-section that we consider is $\sigma_{\chi}\sim \mathcal{O}(10^{-10})\,\mathrm{cm}^2$. 
This leads to a mean free path of dark matter in seawater of $\lambda_{\chi}\simeq 10^{-15}\,\mathrm{m} \times (10^{-10}\,\mathrm{cm}^2)/\sigma_{\chi}$,
where we used that the density of seawater is $\rho_{\rm{sea}}=1.03\,\mathrm{g}/\mathrm{cm}^{3}$.
Such a large cross-section and short mean free path can only be achieved with composite dark matter candidates~\cite{Digman:2019wdm}.
For reference, this distance is considerably smaller than the intermolecular distance in seawater ($\sim 3 \times 10^{-10}\,\mathrm{m}$).

We focus on the regime in which $\mDMeq \gg m_A$, where $m_A$ is the mass of a hydrogen or oxygen nucleus, and where the number of scatters $N$ is large, such that $N \gg 1$. 
In this regime, each collision 
transfers a recoil energy of approximately $E_A\sim (m_A/m_{\chi})\,E_{\chi} \sim \mathcal{O}(10\,\mathrm{keV})$, where $E_\chi =\frac{1}{2}m_\chi v_\chi^2$ is the dark matter kinetic energy.  The recoil is equally likely in any direction,
while the dark matter is deflected by an angle $\delta \theta \sim m_A / (\mDMeq \sqrt{N}) \ll 1$ after $N$ scatters.
Given the large number of scatters and the small deflection angle,
we can model the dark matter's energy loss as a continuous process along a straight path,
with the kinetic energy loss per unit path length given~by
\begin{equation} \label{eq:dEdz}
\frac{\mathrm{d}E_\chi}{\mathrm{d}z} = -\rho_{\mathrm{sea}} \sigma_{\chi} v_{\chi}^2 \exp\left(-\frac{z}{\ell_{\mathrm{sea}}}\right)\,,
\end{equation}
where \(z\) is the distance travelled through the seawater and 
\begin{align} \label{eq:lsea}
\ell_{\mathrm{sea}}&=\frac{m_{\chi}}{2 \rho_{\mathrm{sea}}\sigma_{\chi}}\\
&\simeq 485\,\mathrm{km} \times \left(\frac{m_{\chi}}{10^{-2}\,\mathrm{g}}\right) \left(\frac{10^{-10}\,\mathrm{cm}^2}{\sigma_{\chi}} \right)
\end{align}
 defines the length scale over which the dark matter loses a significant fraction of its kinetic energy (see app.~\ref{App:length} for a derivation).
Given that the Mediterranean Sea is typically only a few kilometres deep, over most of the parameter space that we consider, the dark matter traverses the full seawater depth with approximately constant energy loss per unit length (i.e., $\ell_{\rm{sea}}\gg \mathrm{km}$). 
Moreover, in this regime, $E_\chi$ remains approximately constant since the dark matter loses a negligible fraction of its kinetic energy in each collision.
We focus on this regime in the main text, while app.~\ref{App:lines} addresses the case in which the energy loss varies as dark matter passes through the seawater.

With these considerations, we model
the energy deposition density along the dark matter's path~as
\begin{equation}\label{eq:q_definition}
q(\bm{r}) =\sum_{A={\rm{\{H,O\}}}} \frac{1}{2\pi} \frac{dE_A}{dz} \frac{1}{\sigma_A^2} \exp\left(-\frac{\rho^2}{2 \sigma_{A}^2}\right)\;,
\end{equation}
where we sum over contributions from scattering with hydrogen and oxygen nuclei. Here, $\rho$ is the radial (perpendicular) distance from the energy deposition track centre, such that $|\bm{r}|=\sqrt{\rho^2+z^2}$. The species-dependent terms are~$dE_A/dz$, the energy deposition per unit length, and~$\sigma_A$, the track width parameter.
The approximation of a Gaussian distribution with radial parameter~$\sigma_A$ enables analytic solutions while capturing the essential physics. 
The cylindrical symmetry around the $z$-axis, which is chosen to align with the dark matter's path, follows from the isotropic recoil distribution.
Unless stated otherwise, in our calculations, we use $dE_A/dz = \langle E\rangle_A / \lambda_A$, where $\langle E\rangle_A$  and $\lambda_A$ are the mean nuclear recoil energy and mean free path between collisions for species~$A$, respectively. In the regime $\ell_{\rm{sea}} \gg \mathrm{km}$, since $E_\chi$ is approximately constant along the path, it follows that $\langle E\rangle_A$ and $dE_A/dz$ are also approximately constant. 
In seawater, $dE_{\mathrm{O}}/dz \approx 7.9\,dE_{\mathrm{H}}/dz$,
since the energy transfer during collisions is more efficient for heavier elements.
A derivation of eq.~\eqref{eq:q_definition} and further justifications for the approximations used are given in app.~\ref{App:length}. The main features of the energy deposition are illustrated in fig.~\ref{fig:idea}.

The radial parameter~$\sigma_A$ is
set by the larger of two characteristic length scales. The first is the typical recoil distance $\sigma_{\rm{recoil}}$ of nuclei after a collision with dark matter, which as detailed in app.~\ref{App:length} is $0.14\,\upmu\mathrm{m}$ for oxygen recoils and $0.082\,\upmu\mathrm{m}$ for hydrogen recoils. The second is the radius $R_\chi$ of the ultra-heavy dark matter, which we assume to be composite. Following macro dark matter models \cite{jacobsMacroDarkMatter2015}, we use the geometric approximation for the cross-section, $\sigma_{\chi} \approx 4 \pi R_\chi^2$, such that $ R_\chi \simeq 0.028~\upmu\mathrm{m} \times \sqrt{ (\sigma_{\chi}/10^{-10}\,\mathrm{cm}^2) }$.

\subsection{Attenuation-free solutions}

Having established~$q(\bm{r},t)$ for ultra-heavy dark matter scattering in seawater, we proceed to solve eq.~\eqref{pressure_equation_full} to find the acoustic pressure.
The form of~$q(\bm{r},t)$ is a sum of independent contributions from scattering with oxygen and hydrogen, allowing us to similarly decompose the solution to eq.~\eqref{pressure_equation_full} into these separate components.

In the regime where~$\sigma_A\ll \rho$ --- a condition that holds for any physically relevant case --- the solution for each species $A$ is given by
\begin{equation}\label{eq:p_Afull}
p_A(\rho,t) = \frac{ \alpha}{2\pi c_p }  \frac{dE_A}{dz} \frac{c_s^2}{\sqrt{2 \pi }} \frac{\sigma_A^{-3/2}}{\sqrt{\rho}} \,I_p\left(\frac{t-\rho/c_s}{\sigma_A/c_s}\right)\;.
\end{equation}
The $1/\sqrt{\rho}$ dependence in the amplitude reflects the cylindrical symmetry of the outward-propagating pressure waves, as illustrated in fig.~\ref{fig:idea}. The radial distance~$\rho$ is the only relevant geometric parameter. 
The temporal evolution of the pressure wave is governed by the {\it dimensionless bipolar-pulse function}, $I_p(A)$, which takes $\mathcal{O}(1)$ values, so the overall signal amplitude is determined by the other pre-factors.

\begin{figure}
    \centering
    \includegraphics[width=0.99\linewidth]{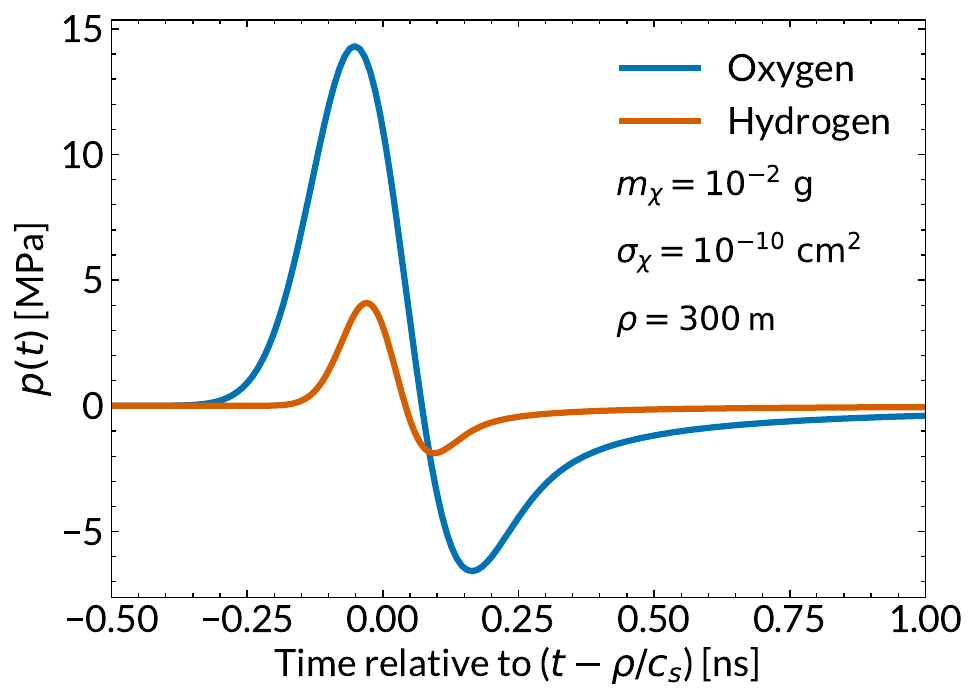}
    \caption{Expected acoustic pressure \textit{without attenuation} from ultra-heavy dark matter with $m_\chi = 10^{-2}\,\mathrm{g}$ and $\sigma_\chi = 10^{-10}\,\mathrm{cm}^2$ at a radial distance $\rho = 300\,\mathrm{m}$.
    We show contributions from oxygen (blue) and hydrogen (red) separately. The full solution is the sum of these signals. The oxygen signal has a greater amplitude and is broader because~$dE_A/dz$ and~$\sigma_A$ are both larger, relative to hydrogen.
    }
    \label{fig:PressureSourceTime}
\end{figure}
 
  The full expression for $I_p(A)$ in terms of modified Bessel functions is derived in app.~\ref{sec:pressure_derivations}.
 For physical insight, we examine its behaviour across three time regimes: early, intermediate, and late times.
 Since at a radial distance $\rho$, the signal arrives after a delay of approximately $\rho/c_s$, this corresponds to $t\ll \rho/c_s$, $t\sim \rho/c_s$ and $t\gg \rho/c_s$.
 
The full bipolar-pulse profile is shown in fig.~\ref{fig:PressureSourceTime}.
We separately plot the solutions for oxygen (blue) and hydrogen (red) at a fixed distance of $\rho=300\,\mathrm{m}$, 
for $m_{\chi}=10^{-2}\,\mathrm{g}$ and $\sigma_{\chi}=10^{-10}\,\mathrm{cm}^2$, values typical of the dark matter parameter regions we aim to explore.
The time is shown relative to the arrival delay $\rho/c_s$, and the full solution is a sum of these two signals.

As the pulse begins to arrive, the pressure follows
\begin{equation}
I_{p}( t\ll \rho/c_s) \approx \sqrt{\frac{\pi}{2} \left| \frac{t-\rho/c_s}{\sigma_A/c_s}\right|} \exp\left(-\frac{(t - \rho/c_s)^2}{2 \sigma^2_A/c_s^2} \right)\,,
\label{eq:pulse_negative_t}
\end{equation}
which shows a signal rising as it sweeps through the tail of the Gaussian profile, reflecting the initial energy deposition profile given in eq.~\eqref{eq:q_definition}. 

At intermediate times, the pulse exhibits a characteristic compression phase followed by a decompression phase, resulting in a bipolar structure,
with a characteristic width governed by $\sigma_A/c_s$.
This bipolar structure mirrors that observed in acoustic signals from UHE neutrinos~\cite{lahmann2011ultra}.

At late times, the asymptotic limit gives
\begin{equation}
I_{p}( t\gg \rho/c_s) \approx - \frac{\sqrt{\pi}}{2}\left(\frac{\sigma_A/c_s}{(t-\rho/c_s)}\right)^{3/2}\;.
\label{eq:pulse_positive_t}
\end{equation}
When substituted into eq.~\eqref{eq:p_Afull}, we obtain a power-law in time that is independent of $\sigma_A$. 
This late-time behaviour arises from the coherent sum of acoustic contributions along the dark matter track,
and is the scaling found for a long, zero-width line-source track~\cite{learnedAcousticRadiationCharged1979}.

\begin{figure}
    \centering
    \includegraphics[width=0.99\linewidth]{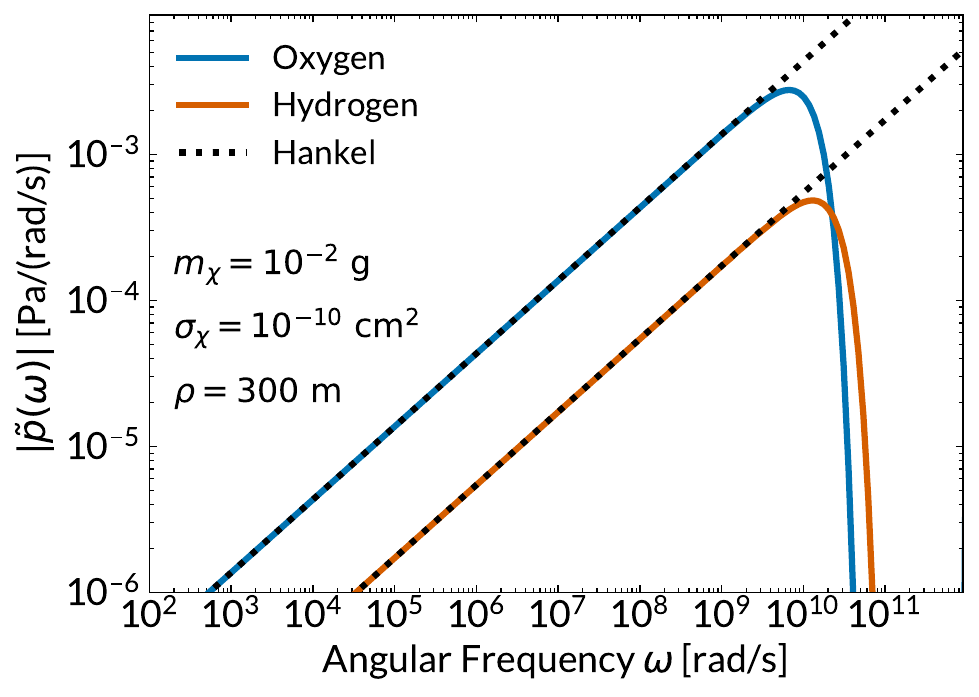}
    \caption{Absolute magnitude of frequency-domain acoustic pressure solutions \textit{without attenuation} from ultra-heavy dark matter with 
    $m_\chi = 10^{-2}\,\mathrm{g}$ and $\sigma_\chi = 10^{-10} \, \mathrm{cm}^2$ at a radial distance $\rho = 300\,\mathrm{m}$. We show contributions from oxygen (blue) and hydrogen (red) separately.
    Dotted lines labelled `Hankel' show an analytic approximation, which provides a good match for $\omega\ll c_s/\sigma_A$.
    }
    \label{fig:PressureSourceFreq}
\end{figure}

 Importantly, the solution in eq.~\eqref{eq:p_Afull} does not account for attenuation effects.
To incorporate attenuation, which we discuss in the next section, we need to understand the frequency spectrum of the pressure pulse.
The frequency-domain representation of the pressure signal, $\tilde{p}(\rho,\omega)$, is obtained by taking the Fourier transform of the time-domain solution. 
As derived in app.~\ref{sec:pressure_derivations}, the solution for each component  takes the form: 
\begin{equation}    
\begin{split}\label{eq:p_freq_noatten}
\tilde{p}_A(\rho,\omega) &= \frac{i \omega \alpha}{2\pi c_p} \frac{dE_A}{dz} 
\int_{0}^{\pi/2} d\theta \, \exp\left( \frac{-i \rho \omega}{c_s} \sec\theta \right) \\
&\qquad \qquad \quad \times \sec\theta  \exp\left( - \frac{\sigma_A^2 \omega^2}{2 c_s^2} \sec^2\theta \right),
\end{split}
\end{equation}
where~$\omega$ is the angular frequency (here, restricted to positive values), $\theta$ is an integration variable, and the complete solution is the sum of the oxygen and hydrogen components.

The frequency-domain solutions are complex-valued, so in fig.~\ref{fig:PressureSourceFreq} we plot the absolute magnitude of $\tilde{p}_A(\rho,\omega)$ for oxygen (red) and hydrogen (blue) separately, at a radial distance of $\rho=300\,\mathrm{m}$ and for
 $m_{\chi}=10^{-2}\,\mathrm{g}$ and $\sigma_{\chi}=10^{-10}\,\mathrm{cm}^2$. 
Both pressure pulses exhibit a broadband frequency spectrum, with a peak at approximately~$10^{10}\,\mathrm{rad/s}$. This peak frequency corresponds to a characteristic scale set by $\omega\sim c_s/\sigma_A$, 
while the broad profile arises as a consequence of the power-law at times $t\gg \rho/c_s$ of the time-domain pressure pulse.
 The oxygen signal has a slightly lower peak frequency than hydrogen due to its larger track width, $\sigma_{\mathrm{O}} > \sigma_{\mathrm{H}}$, but a greater amplitude owing to its larger $dE_A/dz$. In the regime where $\ell_{\rm{sea}}\gg\mathrm{km}$, as assumed here, the pulse properties are independent of $m_{\chi}$ since $dE_A/dz$ is independent of the dark matter mass.

\subsection{Behaviour in the observational frequency range}

The hydrophones proposed for a large underwater array operate in the approximate {\emph{angular}} frequency range $\omega \sim 5\times 10^4\,\mathrm{rad/s}$ to~$5\times 10^5\,\mathrm{rad/s}$~\cite{naumann2007development,buisFibreLaserHydrophones2014, buisFiberOpticHydrophones2016}.
This corresponds to the regime where~$\omega\ll c_s/\sigma_A$, even for $\sigma_A$ values as large as $\mathcal{O}(1\,\mathrm{mm})$. For all parameter values considered in this work, this condition holds.
In the limit $\omega\ll c_s/\sigma_A$, we can expand the $\sigma_A$-dependent exponential term in eq.~\eqref{eq:p_freq_noatten} to leading order, obtaining the analytic expression:
\begin{equation}\label{eq:HankelApproximation}
\tilde{p}_A(\rho,\omega) \approx
\frac{\omega \alpha}{2 \pi c_p }  \frac{dE_A}{dz} \frac{\pi}{2} \, H_0^{(2)}\!\left( \frac{\rho\, \omega}{c_s}\right) \;.
\end{equation}
Here, $H_0^{(2)}(z)$ is the Hankel function of the second kind.

The absolute magnitude of eq.~\eqref{eq:HankelApproximation} is shown as the black dotted lines in fig.~\ref{fig:PressureSourceFreq} for both oxygen and hydrogen. As anticipated, in the typical operational range of the hydrophones, the agreement with the exact result is excellent when $\omega\ll c_s/\sigma_A$. Importantly, this result demonstrates that the specific value of $\sigma_A$ does not affect the solution in the observational range, provided the condition $\omega\ll c_s/\sigma_A$ is satisfied.

We can further simplify the expression for $\tilde{p}_A(\rho,\omega)$ by taking the asymptotic limit  of $H_0^{(2)}(z)$, which is valid for $\rho \gg c_s/\omega$ (approximately $1\,\mathrm{m} \times (10^4\,\mathrm{rad/s})/\omega)$.
In this asymptotic regime, we find
\begin{equation}\label{eq:Hankelapprox_asy}
\tilde{p}_A(\rho,\omega) \sim \frac{ \alpha}{2 \pi c_p }  \frac{dE_A}{dz} \sqrt{\frac{ \pi c_s \omega}{2 \rho }} \exp\left(-\frac{i\rho \omega }{c_s}+\frac{i \pi}{4} \right) \;.
\end{equation}
For the parameter values used in fig.~\ref{fig:PressureSourceFreq}, this asymptotic regime applies, which explains the observed behaviour $\left|\tilde{p}_A(\rho,\omega)\right|\propto \sqrt{\omega}$. 

 Finally, there is a straightforward intuitive explanation for these results when both conditions $\omega\ll c_s/\sigma_A$ and $\rho \gg c_s/\omega$ hold.
 In the time domain, the solution can be expressed~as
 \begin{equation}\label{eq:planewaves}
 \begin{split}
p_A(\rho,t) &\sim \mathrm{Re}\left[\frac{ \alpha}{2 \pi c_p }  \frac{dE_A}{dz} \frac{ c^2_s }{\sqrt{2 \pi  }} \right. \\
&\quad  \times \left. \int_{\rho^{-1}}^{\sigma_A^{-1}}\!dk \sqrt{\frac{k}{\rho }}  \exp\left(i \omega t -ik \rho+\frac{i \pi}{4} \right) \right],
\end{split}
\end{equation}
where $k=\omega/c_s $ and $\mathrm{Re}[\cdot]$ denotes the real part. This shows that the solution is a superposition of cylindrical plane waves propagating outwards from the ultra-heavy dark matter path, which sources the acoustic waves.

\section{Attenuated Dark Matter Induced Thermo-acoustic Radiation}\label{sec:Attenuated}

Acoustic radiation in seawater is attenuated by two main mechanisms: viscous absorption, which occurs in pure water, and chemical relaxation effects due to boric acid and magnesium sulphate -- dissolved salts found in seawater~\cite{francois1982soundI, francois1982soundII}.

The absorption of acoustic radiation is characterised by a frequency-dependent absorption coefficient,~$\tilde{a}(\omega)$. Following refs.~\cite{learnedAcousticRadiationCharged1979, niessUnderwaterAcousticDetection2006}, we parametrise this coefficient~as
\begin{equation}
\tilde{a}(\omega) = \frac{\omega^2}{\omega_0 c_s} + \frac{2}{\lambda_1} \frac{i\omega}{\omega_1 + i\omega} + \frac{2}{\lambda_2} \frac{i\omega}{\omega_2 +i\omega}\;.
\label{eq:alpha}
\end{equation}
The first term represents the absorption by pure water and is characterised by the frequency scale $\omega_0$.
The second and third terms represent chemical relaxation processes from boric acid and magnesium sulphate, respectively. These processes are characterized by distance scales $\lambda_1,\lambda_2$ and frequency scales $\omega_1, \omega_2$. In these chemical processes, the compression or rarefaction of the compounds lags behind the applied pressure, which manifests as a complex compressibility~\cite{LiebermannPhysRev.76.1520}. As we will show, this introduces dispersion that delays the propagation of the bipolar pulse.

The frequency and distance parameters in $\tilde{a}(\omega)$ depend on the seawater conditions: temperature, pH, depth and salinity. For our analysis, we assume conditions typical of the Mediterranean Sea, namely $T=13.5\degree \mathrm{C}$, a pH of $8$, a salinity of $38.5$\textperthousand\ and a depth of $1.2\,\mathrm{km}$~\cite{mcdougall2011getting}. With these choices, we match the real part of $\tilde{a}(\omega)$ to the expression in ref.~\cite{francois1982soundII}, obtaining $\omega_0=4.32\times10^{11}\,\mathrm{rad/s}$, $\omega_1=8.37\times 10^3 \,\mathrm{rad/s}$, $\omega_2=5.827\times 10^5\,\mathrm{rad/s}$, $\lambda_1=64.4\,\mathrm{km}$ and $\lambda_2=152.7\,\mathrm{m}$. Note that we use units where $\tilde{a}(\omega)$ is measured in inverse-metres, so we have applied the appropriate conversion factor when matching to the expression in ref.~\cite{francois1982soundII}.\footnote{Our value for $\omega_0$ differs from ref.~\cite{niessUnderwaterAcousticDetection2006} when the same values of temperature, pH, depth and salinity are used. The difference seems to arise from their use of the $P_2$ and $P_3$ expressions from refs.~\cite{francois1982soundI,francois1982soundII}, where pressure and depth appear to have been interchanged (see app.~F of ref.~\cite{niess:tel-00132273}).}

To account for attenuation, it is most straightforward to work in the frequency domain, where each frequency mode experiences attenuation according to
\begin{equation}
    \tilde{p}_{a}(\rho, \omega) = \exp\left(-\frac{ \tilde{a}(\omega)\rho}{2}\right)\, \tilde{p}(\rho, \omega),
    \label{eq:attenuationModel}
\end{equation}
where $\rho$ is the relevant distance scale since each mode propagates outward as a cylindrical wave, and the factor of $1/2$ is included because we are dealing with the pressure amplitude rather than the intensity~\cite{urick1983principles}.
Here, $\tilde{p}(\rho,\omega) = \tilde{p}_{\rm{O}}(\rho,\omega)+\tilde{p}_{\rm{H}}(\rho,\omega)$.
The attenuated signal in the time domain can be recovered via the inverse Fourier transform. Since the signal is real-valued, this is expressed as
\begin{equation}
    p_{a}(\rho, t) =\mathrm{Re}\left[\frac{1}{\pi}\int_0^{\infty} d\omega \,\tilde{p}_a(\rho, \omega) \exp(i \omega t)\right]\;.
    \label{eq:pt_attenuationModel}
\end{equation}

\subsection{Attenuation effects in pure water}

To gain intuition for the impact of attenuation, we first consider the case of pure water (i.e., we set the second and third terms in eq.~\eqref{eq:alpha} to zero). In frequency space, the pure water attenuation introduces a distance-dependent cut-off at the angular frequency scale $\sqrt{\omega_0 c_s/\rho} \approx  1.5\times 10^6 \cdot \sqrt{300\,\mathrm{m}/\rho}  \;\mathrm{rad/s}$. 
This behaviour is illustrated in fig.~\ref{fig:PressureAttenFreq}, where the pure water-attenuated signal is shown as a dotted line for radial distances of $300\,\mathrm{m}$ (blue) and $1\,\mathrm{km}$ (pink).

We can derive an analytic expression for the pure water-attenuated time-domain signal in the limits $\omega\ll c_s/\sigma_A$ and $\rho \gg c_s/\omega$.
Substituting eq.~\eqref{eq:Hankelapprox_asy} into eq.~\eqref{eq:pt_attenuationModel}, we obtain
\begin{equation}\label{eq:pure_pt_attenuated}
\begin{split}
p^{\rm{pure}}_a(\rho,t) &= \frac{\alpha}{2 \pi c_p} \frac{dE}{dz} \frac{c_s^2}{\sqrt{2 \pi}} \frac{1}{\sqrt{\rho}} \\
&\quad \times \left(\sqrt{\frac{\rho c_s}{\omega_0}}\, \right)^{-3/2} I_p\left( \frac{t - \rho/c_s}{\sqrt{\rho c_s / \omega_0}/c_s} \right),
\end{split}
\end{equation}
where $dE/dz=dE_{\mathrm{O}}/dz+dE_{\mathrm{H}}/dz$. 
We have expressed this in form that allows for a straightforward comparison with the unattenuated result.
Upon comparing eqs.~\eqref{eq:p_Afull} and~\eqref{eq:pure_pt_attenuated}, we see that we have the same functional form, but with the replacement $\sigma_A \to \sqrt{\rho c_s / \omega_0}$. 

This change leads to two distance-dependent effects.
Firstly, the attenuated signal amplitude now scales like $\rho^{-5/4}$ instead of $\rho^{-1/2}$. As a result, the attenuated signal amplitude is reduced by a factor $10^6 \cdot \left( \rho/300\,\mathrm{m}\right)^{3/4} \cdot \left( \sigma_A/0.1 \,\upmu \mathrm{m} \right)^{-3/2}$ relative to the unattenuated signal. Secondly, while the pulse shape remains bipolar, as indicated by the appearance of the function $I_p(A)$, the pulse width has broadened significantly by a factor $10^4 \cdot \left( \rho/300\,\mathrm{m}\right)^{1/2} \cdot \left( \sigma_A/0.1 \,\upmu \mathrm{m} \right)^{-1}$ relative to the unattenuated signal. 

Therefore, for the parameters used in fig.~\ref{fig:PressureSourceTime}, which led to a pressure pulse with a $10\,\mathrm{MPa}$-scale amplitude and a bipolar pulse width on the order of $0.1\,\mathrm{ns}$, for the same parameters, we expect an attenuated pressure at $\sim10\,\mathrm{Pa}$ with a pulse duration on the~$\upmu\mathrm{s}$ timescale. 

\subsection{Attenuation effects in seawater}

\begin{figure}
    \centering
    \includegraphics[width=0.99\linewidth]{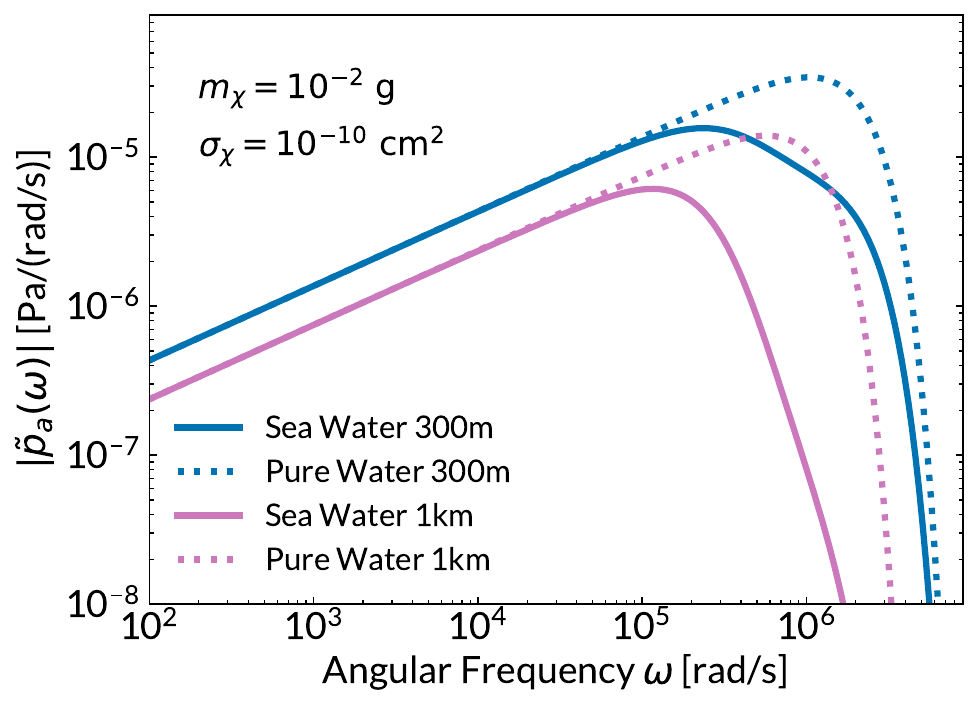}
    \caption{Absolute magnitude of frequency-domain acoustic pressure in seawater (solid lines) and pure water (dotted lines) from ultra-heavy dark matter with $m_\chi = 10^{-2}\,\mathrm{g}$ and $\sigma_\chi = 10^{-10} \, \mathrm{cm}^2$ at radial distances of $\rho = 300\,\mathrm{m}$ (blue) and $\rho = 1\,\mathrm{km}$ (pink). 
     The absorption coefficient introduces distance-dependent frequency cut-offs, most prominently in seawater, leading to significant suppression at high frequencies compared to the unattenuated case in fig.~\ref{fig:PressureSourceFreq}.
}
    \label{fig:PressureAttenFreq}
\end{figure}

We now consider the case of seawater, where the full expression for the absorption coefficient is used. Each of the three terms in eq.~\eqref{eq:alpha} in principle leads to a distance-dependent cut-off in the frequency spectrum. In practice, since we only consider radial distances up to a few~km, the boric acid term has a negligible impact as we work in the regime where $\rho\ll \lambda_1$. For magnesium sulphate, there is an angular frequency cut-off at approximately $\sqrt{\omega_2^2 \lambda_2/(2\rho)}\approx 3\times 10^5 \cdot \sqrt{300\,\mathrm{m}/\rho}  \;\mathrm{rad/s} $, which is approximately five times smaller than the cut-off in pure water. This behaviour is evident when comparing the pure water (dotted lines) and seawater results (solid lines) shown in fig.~\ref{fig:PressureAttenFreq}. The seawater spectrum exhibits an additional feature at $\rho=300\,\mathrm{m}$, where a distinctive shoulder near $5\times10^5\,\mathrm{rad/s}$ is observed. This arises because the real part of the magnesium sulphate absorption coefficient approaches an asymptotic value of $1/\lambda_2$ at high frequencies, so the exponent in eq.~\eqref{eq:attenuationModel} tends to $-\rho/\lambda_2$. At $\rho=1\,\mathrm{km}$, where the distance significantly exceeds $\lambda_2$, this term becomes negligible, resulting in a frequency response that more closely resembles the case of pure water, but with a lower frequency cut-off.

\begin{figure}
    \centering
    \includegraphics[width=0.99\linewidth]{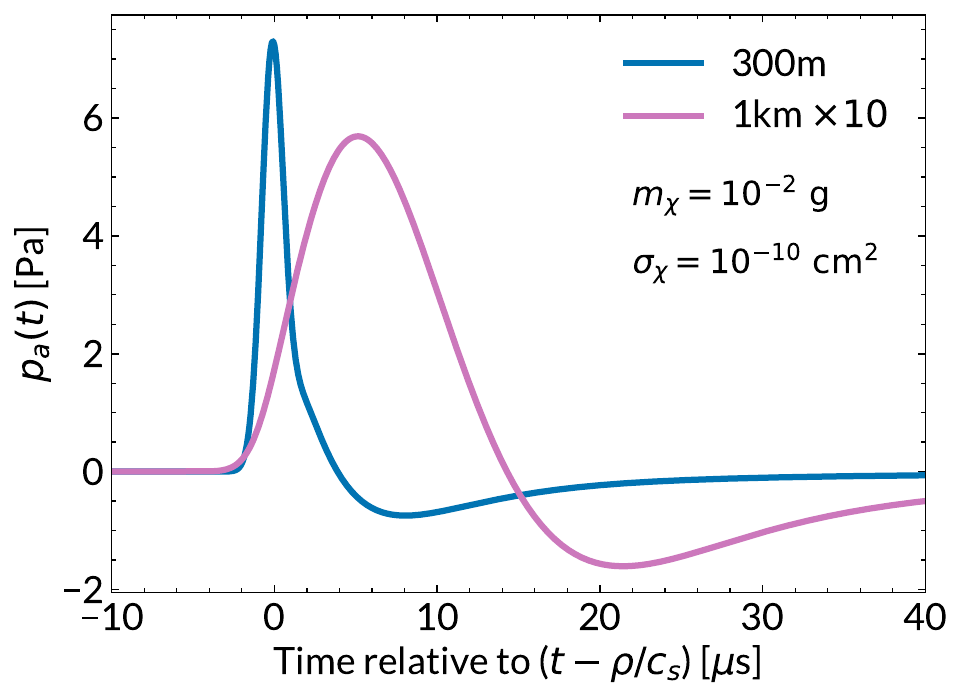} 
    \caption{
     Expected acoustic pressure in seawater from ultra-heavy dark matter with $m_\chi = 10^{-2}\,\mathrm{g}$ and $\sigma_\chi = 10^{-10}\,\mathrm{cm}^2$ at radial distances of $\rho = 300\,\mathrm{m}$ (blue) and $\rho = 1\,\mathrm{km}$ (pink, multiplied by $10$ for clarity). Compared to the unattenuated pulse in fig.~\ref{fig:PressureSourceTime}, attenuation causes significant amplitude reduction and temporal broadening. The signal experiences a small delay in arrival time, which is most pronounced for the $\rho = 1\,\mathrm{km}$ pulse, due to dispersion from the absorption coefficient's imaginary component.
    }
   \label{fig:PressureAttenTime}
\end{figure}

Next, we examine the seawater attenuated signal in the time domain. The results are shown in fig.~\ref{fig:PressureAttenTime} for $\rho = 300\,\mathrm{m}$ (blue) and $1\,\mathrm{km}$ (pink)
for $m_{\chi}=10^{-2}\,\mathrm{g}$ and $\sigma_{\chi}=10^{-10}\,\mathrm{cm}^2$, where the $1\,\mathrm{km}$ pulse has been rescaled by a factor of~10 for clarity.  The frequency-dependent distortions observed in fig.~\ref{fig:PressureAttenFreq} manifest as modifications to the bipolar pulse structure. Three distinct phenomena emerge relative to the pure water scenario.  

Firstly, we observe a delay in the arrival time of the pulse. This is most evident for $\rho=1\,\mathrm{km}$ where the pulse is shifted by approximately $15\,\upmu s$. This delay in the arrival time arises from the imaginary component of $\tilde{a}(\omega)$, which acts as an effective frequency-dependent refractive index that modifies the group velocity of the propagating wave. Quantitatively, the speed of sound is reduced by $\Delta c_s\approx -c_s^2 \partial (\mathrm{Im}~\tilde{a})/\partial\omega\approx -c_s^2/(\lambda_2 \omega_2)\approx -2.6\,\mathrm{cm}/\mathrm{s}$.

\begin{figure}
    \centering
    \includegraphics[width=0.95\linewidth]{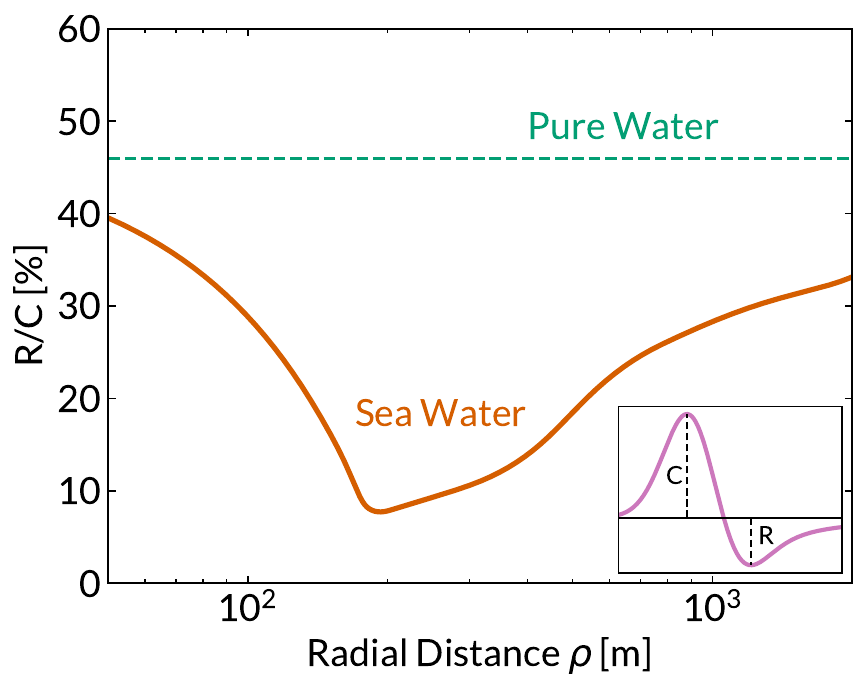}
    \caption{
    The ratio of the height of the rarefaction peak (R) and the compression peak (C) as a function of distance.  The asymmetry in seawater arises from the magnesium sulphate absorption dominating at this distance scale, with the minimum occurring around $\rho\sim\lambda_2$. In contrast, the pure water case maintains a constant R/C ratio.
    The inset illustrates the definition of the R and C peaks in the pressure-time domain.}
    \label{fig:RoverC}
\end{figure}

The second modification is to the pulse asymmetry, which we characterise by the ratio of the minimum-to-maximum amplitudes of the pulse, also known as the rarefaction (R) and compression (C) peaks, respectively. In fig.~\ref{fig:PressureAttenTime}, we see that the asymmetry is more pronounced at $300\,\mathrm{m}$ compared to the $1\,\mathrm{km}$ pulse. More generally, the R/C ratio as a function of radial distance is shown in fig.~\ref{fig:RoverC}. We see that the seawater pulses have a smaller R/C ratio than pure water for all distances. The minimum at just below $200\,\mathrm{m}$ is determined by the length scale $\lambda_2$.

Finally, the bipolar pulse scaling in seawater is steeper than both the unattenuated case ($\rho^{-0.5}$) and pure water case ($\rho^{-1.25}$). In seawater, the scaling follows three distinct regimes: approximately $\rho^{-1.36}$ below 50\,m, $\rho^{-1.92}$ between 50\,m and 500\,m, and $\rho^{-1.43}$ at larger distances.

\section{Sensitivity Analysis \\and Detection Prospects}\label{sec:projections}

Having characterised the properties of the pressure pulse generated by ultra-heavy dark matter traversing seawater, we now turn to the experimental prospects for detection using large-scale hydrophone arrays. 
Early feasibility studies explored the sensitivity of existing underwater arrays to acoustic UHE neutrino detection~\cite{Lehtinen:2001km}. 
The main challenge in detecting ultra-heavy dark matter arises from its extremely low flux, which requires detectors with exceptionally large volumes.
Additionally, we must characterise the detection efficiency, which depends on both the signal detection threshold, $p_{\mathrm{thr}}$, determined by experimental noise sources and the sensitivity of the hydrophone, and the maximum pressure amplitude induced by dark matter at a given distance, $p_{\mathrm{max}}(\rho) = \mathrm{max}_t[\,p_a(\rho,t)\,]$.

For a dark matter candidate, the number of detectable events is given by
\begin{equation}
\begin{split}
N_{\mathrm{events}}= \phi_{\chi} (m_{\chi}) \cdot \text{A}_{\rm{array}} \cdot \eta\left(\frac{d E}{d z}\left(\sigma_\chi, v_\chi\right); p_{\mathrm{thr}}\right),
\label{eq:Nevents}
\end{split}
\end{equation}
where $\phi_\chi$ is the dark matter flux, $\text{A}_{\rm{array}}$ is the effective area of the hydrophone array, and $\eta$ is the detection efficiency, defined as the fraction of dark matter tracks within the array volume that produces an observable signal.\footnote{This treatment assumes the average dark matter speed and energy deposition rate, $dE/dz$, which is sufficient for our sensitivity estimates. A more sophisticated approach would integrate over the dark matter velocity distribution, but this is beyond the scope of this work.}

In what follows, we examine each component that determines the sensitivity.
We then present sensitivity projections for a hypothetical large-scale hydrophone array.

\subsection{Dark matter flux}
\label{subsec:flux}

\begin{figure}
    \centering
    \includegraphics[width=0.99\linewidth]{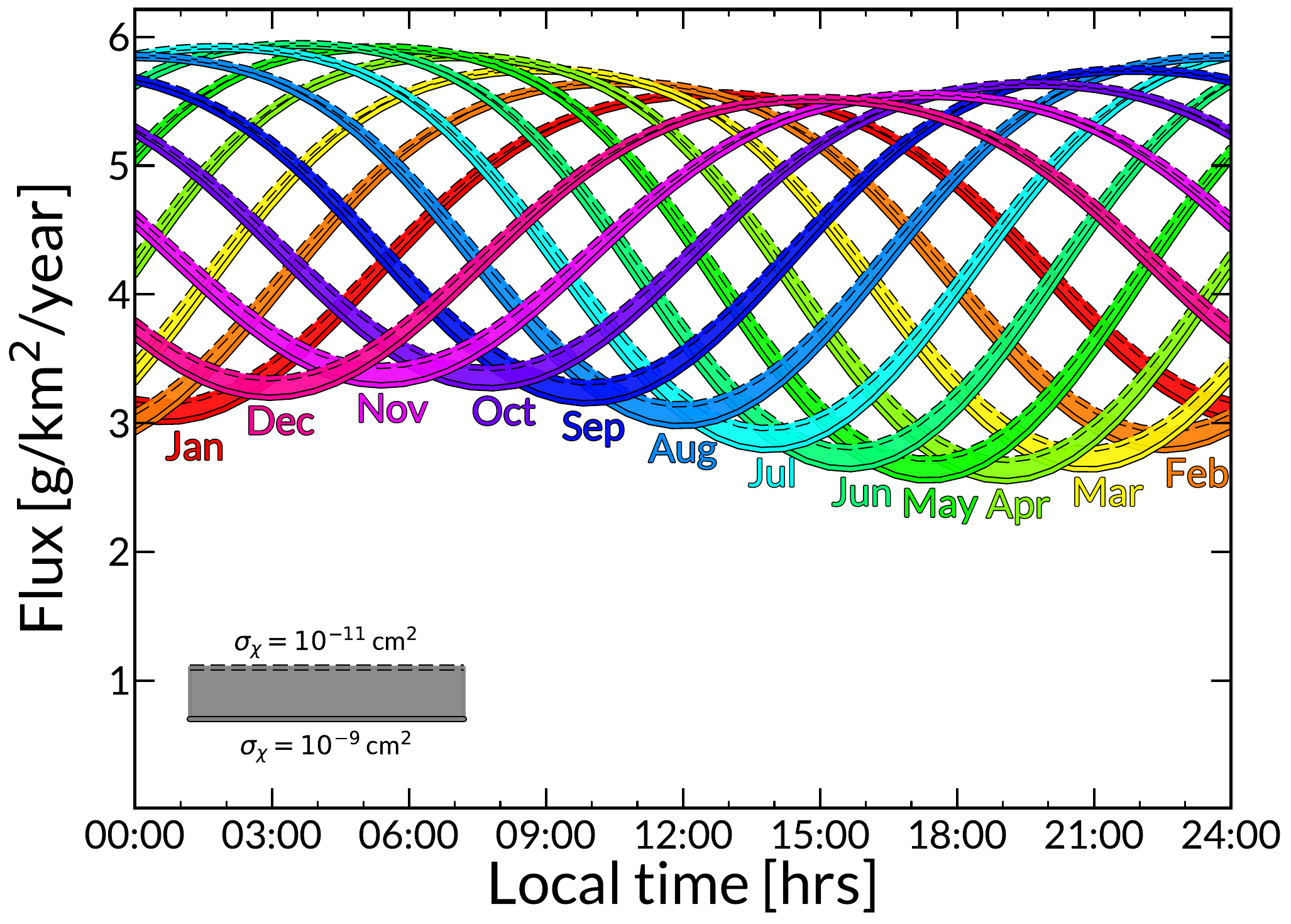}
    \includegraphics[width=0.99\linewidth]{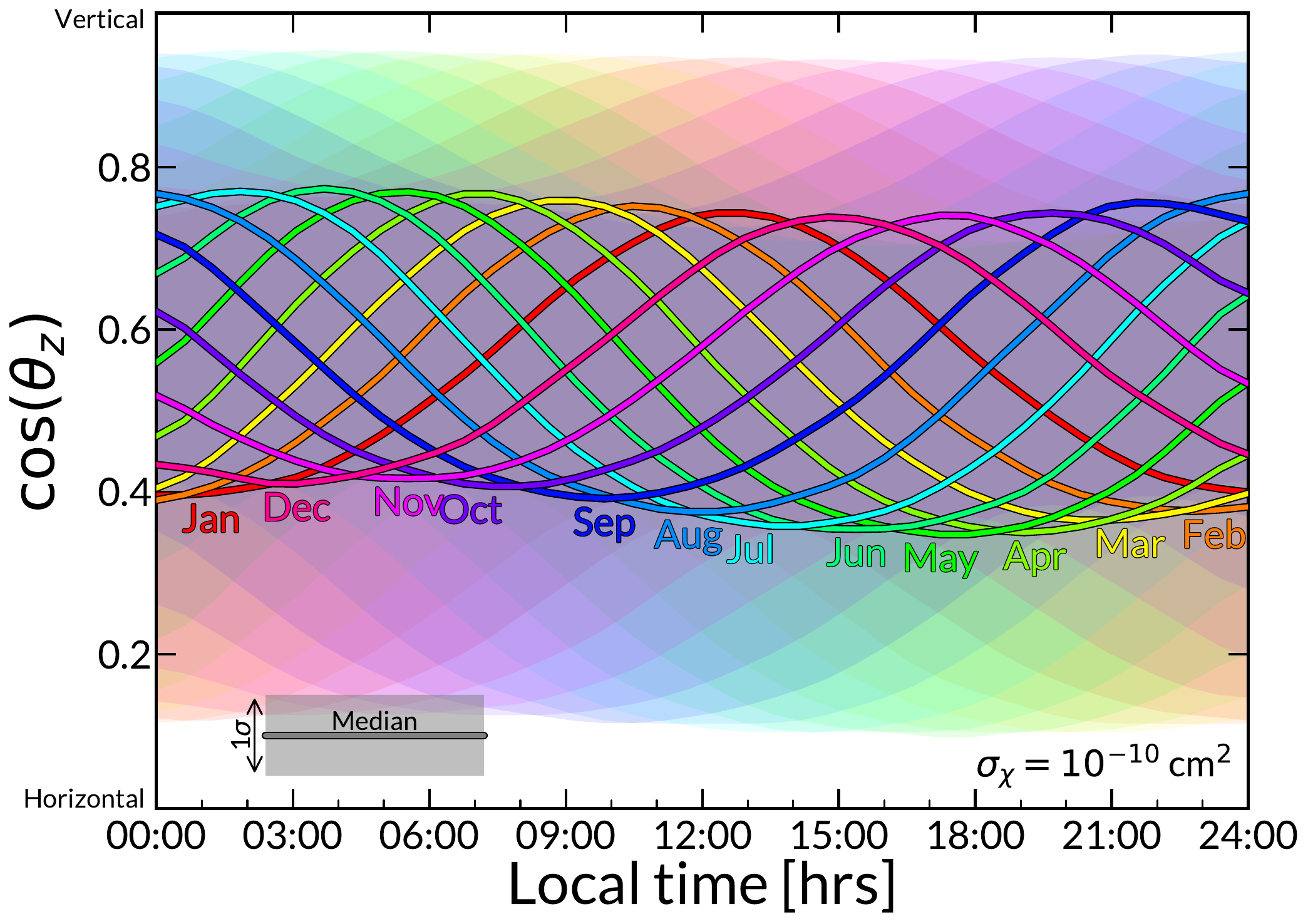}
    \caption{
     Upper panel: Daily modulation of the ultra-heavy dark matter flux at the Mediterranean Sea location, showing a $\sim$50\% modulation amplitude caused by the Earth blocking particles from below. 
 Results are shown for $\sigma_{\chi} = 10^{-9}$~cm$^2$ (solid line) and $10^{-11}$~cm$^2$ (dashed line).
Lower panel: Daily modulation of ultra-heavy dark matter track angle relative to vertical. Solid lines show the median $\cos{\theta_z}$, with shaded regions enclosing 68\% of the distribution around the median. Modulations  are calculated from the first day of each month.}
    \label{fig:FluxModulation}
\end{figure}

Under the assumptions of the Standard Halo Model, with a local dark-matter density of $0.3\,\mathrm{GeV}/\mathrm{cm}^3$, the flux of dark matter particles reaching the top of the Earth's atmosphere is predicted to be $\phi_\chi \approx 6 \left(1 \mathrm{g}/m_{\chi}\right) \mathrm{km}^{-2}\, \mathrm{yr}^{-1}$.
However, this atmospheric flux differs from the flux through the detector array due to two effects.
Firstly, dark matter particles arriving from below the detector must traverse the Earth's interior. For large scattering cross-sections, these particles may be stopped within the Earth, thereby reducing the flux from those directions. Secondly, the dark matter flux will exhibit a distinctive daily modulation. 
This modulation occurs because, while dark matter particles arrive from a range of directions due to the velocity dispersion in the galactic halo, the solar motion through the halo induces a preferred arrival direction aligned with the constellation Cygnus~\cite{Vahsen:2021gnb}.
As the Earth rotates on its axis, the hydrophone array  continuously changes its orientation relative to this directional dark matter flux.

We illustrate this daily modulation in the upper panel of fig.~\ref{fig:FluxModulation}, where we plot the expected flux as a function of local time at the location of the Mediterranean Sea. Our calculation assumes the Standard Halo Model for the velocity distribution~\cite{Evans:2018bqy,Baxter:2021pqo} and incorporates the laboratory velocity relative to the galactic centre~\cite{McCabe:2013kea,Mayet:2016zxu}. At this location, the flux varies between approximately $3-6 \left(1 \mathrm{g}/m_{\chi}\right) \mathrm{km}^{-2}\, \mathrm{yr}^{-1}$, with a mean value of $\phi_\chi \sim 4 \left(1 \mathrm{g}/m_{\chi}\right) \mathrm{km}^{-2}\, \mathrm{yr}^{-1}$. The flux reaches its maximum when the Cygnus constellation is directly above the hydrophone array and its minimum when the constellation is on the opposite side of the Earth, where incoming particles are blocked by the Earth. For larger cross-sections, we observe an additional suppression of the flux. This arises because particles with shallow zenith angles are stopped entirely by the Earth's atmosphere, leading to a slightly reduced flux for $\sigma_{\chi} = 10^{-9}~\mathrm{cm}^2$ (solid line) compared to $10^{-11}~\mathrm{cm}^2$ (dashed line).

The lower panel of fig.~\ref{fig:FluxModulation} shows the corresponding modulation in the zenith angle distribution. While this distribution confirms that all particles arrive from above the detector, the broad distribution of incident dark matter directions makes this a less distinctive signature than the flux modulation.

We also see from both panels in fig.~\ref{fig:FluxModulation}  that the phase of this modulation shifts by exactly 24 hours over the course of a year. This occurs because the dark matter flux follows the sidereal rather than the solar day. If detected, this characteristic temporal pattern, combined with the $\sim$10\% annual modulation in the flux amplitude, would provide an unambiguous signature of galactic origin. 

\subsection{Array configuration}\label{subsec:array_config}

The detection of ultra-heavy dark matter candidates with masses exceeding $1\,\upmu\mathrm{g}$ requires arrays with kilometre-scale dimensions to achieve meaningful event rates. This requirement aligns with proposals for UHE neutrino detection, where similar considerations have led to designs with hydrophones distributed throughout volumes of $\mathcal{O}(100\,\mathrm{km^3})$~\cite{Lahmann:2019unc}. 

For our analysis, we consider a detector array in the Mediterranean Sea with horizontal dimensions of $10\,\mathrm{km} \times 10\,\mathrm{km}$ and a vertical depth of $1\,\mathrm{km}$, positioned with its top at a depth of $1.2\,\mathrm{km}$, consistent with the attenuation parameters assumed in secs.~\ref{sec:Acoustic_formalism} and~\ref{sec:Attenuated}. The array consists of a regular $45 \times 45 \times 10$ grid of hydrophones, spaced uniformly throughout the volume to achieve a density of approximately 200 hydrophones per km$^3$. This density lies at the lower end of the range considered in previous acoustic UHE neutrino studies~\cite{Karg:2006mv,VanOers2023}.

While the physical array covers an area of $10\,\mathrm{km} \times 10\,\mathrm{km}$, in our efficiency calculations, we take the effective area to be $A_{\rm{array}} = 10.5\,\mathrm{km} \times 10.5\,\mathrm{km}$ to account for edge effects and tracks originating outside the nominal array boundaries. We have verified that small variations in this area (e.g., extending to $11\,\mathrm{km} \times 11\,\mathrm{km}$) yield negligible differences in our results. Significantly larger extensions would require modelling the effects on the pressure pulse of refraction and reflection from the surface and seabed~\cite{niessUnderwaterAcousticDetection2006,lahmann2011ultra}, which lie beyond the scope of this study.

\subsection{Background considerations}
\label{subsec:backgrounds}

A characterisation of the noise sources that affect the underwater array is needed to determine the detectability of ultra-heavy dark matter-induced pressure pulses. Since hydrophones are typically optimised to operate at specific frequencies, we focus our analysis on noise sources in the $10$--$100$\,kHz frequency band, which corresponds to the typical operating range for acoustic neutrino detection.\footnote{Here, we refer to \emph{frequency} rather than \emph{angular frequency}.}

In this frequency range, the dominant source of ambient noise arises from sea-surface agitation due to weather conditions, characterised by sea states $0$--$9$~\cite{urick1984ambient}. 
Studies of ambient noise at depths relevant for acoustic detection have investigated the spectral properties and stability of these noise sources (see e.g.,~\cite{Kurahashi:2008zz}).
The fibre-optic hydrophone designs proposed for acoustic neutrino experiments aim to achieve levels of self-noise comparable to sea state $0$~\cite{buisFibreLaserHydrophones2014}, ensuring that the experiment will be limited by sea state noise rather than instrumental effects. Measurements from the O$\nu$DE (Ocean noise Detection Experiment), a real-time experiment to monitor acoustic signals in the Mediterranean Sea at a depth of 2\,km, found average levels of acoustic noise of $5.4 \pm 2.2\,\mathrm{(stat)} \pm 0.3\,\mathrm{(sys)}$\,mPa in the $20$ -- $43$\,kHz band, consistent with sea state 2 conditions~\cite{NEMO:2009wot}. 
Similarly, AMADEUS, also installed in the Mediterranean Sea at a depth of about 2.5\,km measured a mean noise level of $10.1^{+3}_{-2}$\,mPa in the $10$ -- $50$\,kHz band~\cite{ANTARES:2011hls}. Since ambient noise levels typically increase toward lower frequencies, the higher value measured by AMADEUS is expected because of the lower frequency window.

The array must also contend with transient noise sources that can exceed the ambient background. For instance, dolphins emit signals with spectral characteristics similar to those expected from both UHE neutrino and dark matter events~\cite{notarbartolo2016marine}, while shipping traffic can also temporarily increase noise levels~\cite{Lahmann:2017hsh}.  However, these biological and anthropogenic sources act as point sources of noise, while ultra-heavy dark matter produces a characteristic cylindrical pressure wave that typically extends over several hundreds of metres. Furthermore, significant progress has already been made in reducing these transient backgrounds through machine learning and clustering algorithms, as demonstrated by the AMADEUS experiment~\cite{neffSignalClassificationAcoustic2012}. In our analysis, we assume an idealised scenario where there is perfect discrimination against transient noise sources away from the surface.

A final consideration comes from UHE neutrino events, which these arrays are primarily designed to detect. These neutrino events are a background to a dark matter search since they produce acoustic signals with a bipolar pulse shape that peaks in the same frequency range to those expected from dark matter. The rate of UHE neutrino events is expected to be a few events per year for a volume of $100\,\mathrm{km^3}$~\cite{Engel:2001hd, Graf:2010me,lahmann2011ultra}.
Neutrino-induced tracks typically extend over $\mathcal{O}(10)\,\mathrm{m}$, in contrast to the much-longer tracks expected from dark matter.\footnote{This refers to hadronic showers. Electromagnetic showers from electron-neutrino charged-current interactions can be longer, but have distinctive multi-peak structures~\cite{EKonishi_1991, misaki2019historical} that distinguish them from both hadronic showers and dark matter signals.} This track-length difference should provide a handle to distinguish ultra-heavy dark matter events from the UHE neutrino events. However, detailed studies of track reconstruction lie beyond the scope of this work.

\subsection{Detection efficiency}

\begin{figure}[t!]
    \centering
    \includegraphics[width=0.99\linewidth]{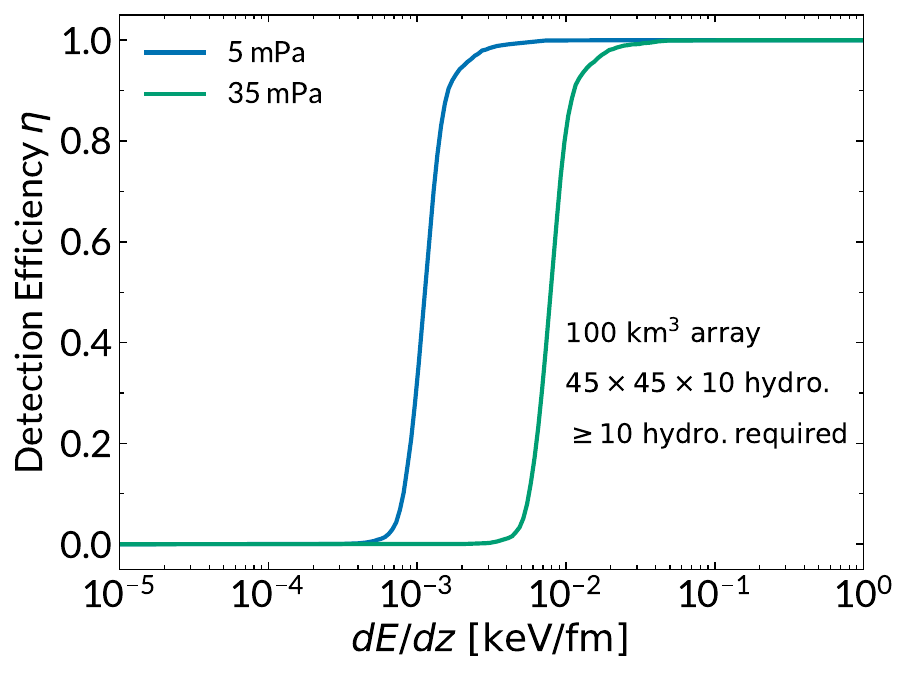}
    \caption{Detection efficiency as a function of energy deposition per unit length for tracks with constant energy deposition, shown for two signal detection threshold values: $5\,\mathrm{mPa}$ (blue) corresponding to an optimistic scenario, and $35\,\mathrm{mPa}$ (green) corresponding to a more conservative} scenario. The detection efficiency assumes a $100\,\mathrm{km}^3$ volume instrumented with $45 \times 45 \times 10 $ hydrophones, where detection of each event requires signals above threshold in at least ten hydrophones.
    
    \label{fig:detection_efficiency}
\end{figure}

We determine the hydrophone array's detection efficiency for tracks with constant energy deposition per unit length through a bespoke Monte Carlo simulation of tracks traversing the array volume. Specifically, we simulate $\mathcal{O}(10^5)$ tracks that intersect a plane of area $10.5 \,\mathrm{km} \times 10.5 \,\mathrm{km}$ through the array's centre. The choice of the array centre for this plane is arbitrary, and we have verified that variations in this position do not lead to appreciable changes in our results. 
The tracks are generated with azimuthal angles distributed uniformly between $0$ and $2\pi$ and zenith angles sampled uniformly between $\cos(\theta_z) = 0.4$ and $0.8$, motivated by the angular distribution shown in fig.~\ref{fig:FluxModulation}. For each track, we calculate the acoustic pressure at every hydrophone position in the $45\times45\times10$ array configuration. Following the reconstruction requirements for cylindrical sources established in ref.~\cite{lahmann2011ultra}, we classify a track as detected if it produces a pressure exceeding the threshold $p_{\mathrm{thr}}$ in at least ten hydrophones. The detection efficiency is given by the fraction of simulated tracks that satisfy this detection criterion.

Figure~\ref{fig:detection_efficiency} shows the detection efficiency as a function of the energy deposition per unit length, $dE/dz$, for two values of $p_{\mathrm{thr}}$: 5\,mPa and 35\,mPa. 
As we showed in fig.~\ref{fig:PressureAttenFreq}, attenuated dark matter-induced acoustic signals are expected to peak in the $10$ -- $50$\,kHz frequency range, similar to neutrino-induced signals.
The value of 5\,mPa corresponds to a detection threshold taken in several UHE neutrino studies~\cite{Karg:2006mv, Graf:2010me, lahmann2011ultra}.
Based on the measured noise levels of $\sim 5$ -- $10$\,mPa in this frequency band, this represents an optimistic threshold that would require multi-detector correlation to achieve sensitivity below the noise level measured in a single hydrophone.
The extended track length of dark matter events and our requirement of a signal above threshold in at least ten hydrophones means that multiple hydrophones will detect the same event, with time delays from the finite speed of sound.
When signals from these detectors are properly time-aligned and summed, the dark matter signal adds coherently while uncorrelated detector noise adds incoherently, thus providing a naive effective $\sqrt{N}$ improvement in signal-to-noise ratio for $N$ triggered detectors.
In contrast, 35\,mPa represents a more conservative threshold 
that ensures a signal-to-noise ratio exceeds unity even for a single hydrophone, without requiring multi-detector correlation.
From fig.~\ref{fig:detection_efficiency}, for the optimistic threshold of 5\,mPa, the detection efficiency exceeds $50\%$ for $dE/dz\gtrsim 9 \times 10^{-4}\,\mathrm{keV}/\mathrm{fm}$. 
Increasing the threshold to 35\,mPa reduces the sensitivity by a factor of seven. This reflects a linear dependence of detection efficiency on pressure threshold that we verified across the range $1$~--~$50$\,mPa.

\subsection{Projected sensitivity reach}

\begin{figure}
    \centering
    \includegraphics[width=0.99\linewidth]{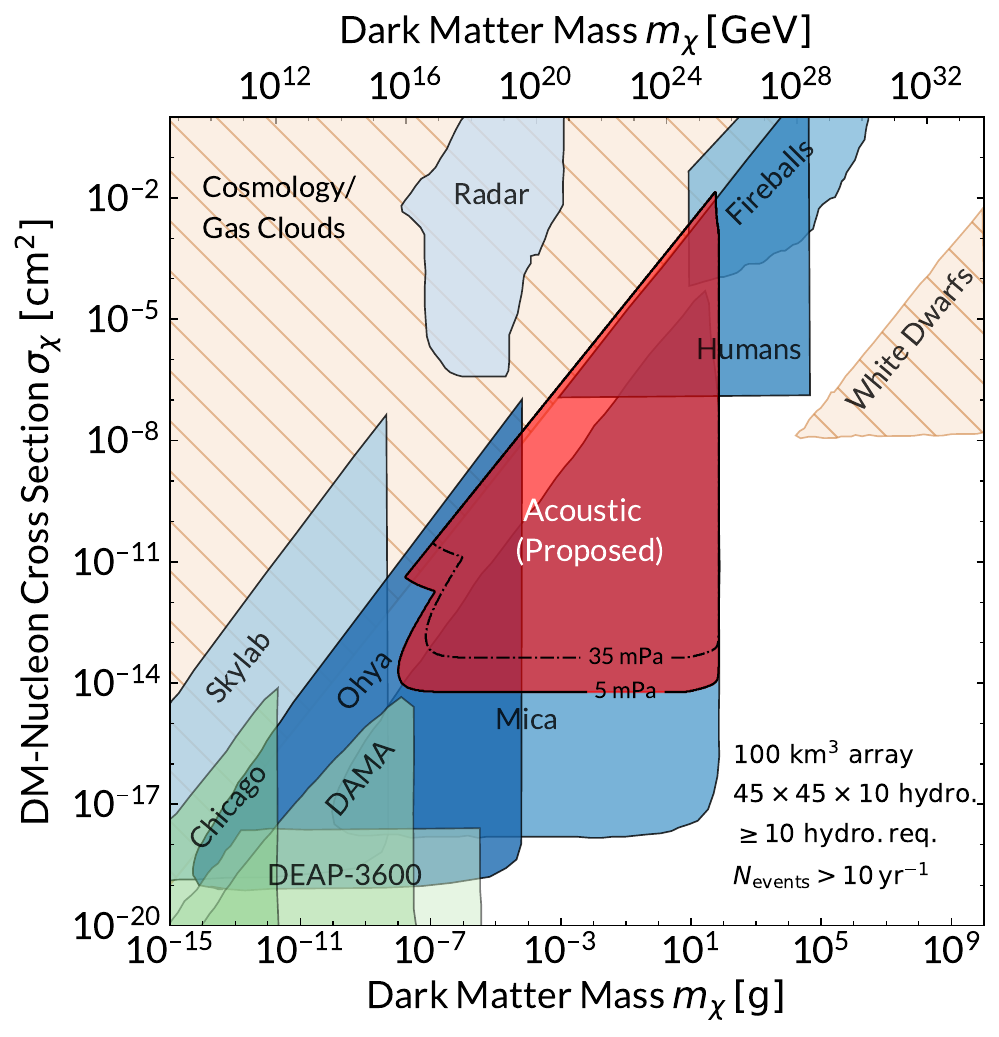}
    \caption{The red filled region shows the projected sensitivity reach in the spin-independent cross-section--dark matter mass plane of a $100\,\mathrm{km^3}$ volume instrumented with $45 \times 45 \times 10 $ hydrophones, assuming $p_{\mathrm{thr}} = 5\,\mathrm{mPa}$ (solid boundary) and $p_{\mathrm{thr}} = 35\,\mathrm{mPa}$ (dot-dashed boundary), signals in at least 10 hydrophones per event, and at least 10 events per year. Light-orange-hatched regions show exclusion regions from cosmology and astrophysics~\cite{Dvorkin:2013cea,grahamWhiteDwarfsDark2018, SinghSidhu:2019tbr,bhoonahDetectingCompositeDark2021, Nadler:2019zrb}, blue from non-traditional search strategies or repurposed experimental data~\cite{PhysRevLett.52.1265, Acevedo:2021tbl,dhakalNewConstraintsmacroscopic2023,sidhumacroscopicDarkMatter2019,sidhuDeathSeriousInjury2020,Niedermann:2024hoa,bhoonahEtchedPlasticSearches2021}, and green from underground terrestrial experiments~\cite{bernabeiExtendedLimitsNeutral1999, cappielloNewExperimentalConstraints2021,adhikariFirstDirectDetection2022}.
    }
    \label{fig:acoustic_contraints}
\end{figure}

Having characterised the array configuration, signal threshold requirements, and detection efficiency, we now establish the sensitivity reach of a hypothetical hydrophone array to ultra-heavy dark matter. As a baseline scenario, we require at least 10 dark matter events per year. While this choice is somewhat arbitrary, it is comparable to rates considered in UHE neutrino studies. The event rate is proportional to the scattering cross-section, so a linear change in the required number of signal events corresponds to a linear change in the sensitivity to $\sigma_{\chi}$.

The characteristic energy-loss length scale in seawater, $\ell_{\rm{sea}}$, determines several boundaries in our sensitivity analysis.  From eq.~\eqref{eq:lsea}, when $\sigma_\chi/m_\chi \gtrsim 2.4 \times 10^{-4}\,\mathrm{cm^2/g}$, we find $\ell_{\mathrm{sea}} \lesssim 20\,\mathrm{m}$. For such short characteristic lengths, energy deposition occurs predominantly near the surface, making it difficult to discriminate against surface-produced background signals.\footnote{This regime would also lead to significant attenuation as dark matter traverses the atmosphere.} Therefore, we restrict our analysis to parameter space where $\ell_{\mathrm{sea}} >20\,\mathrm{m}$.
In contrast, when $\ell_{\mathrm{sea}} \gtrsim 20\,\mathrm{km}$, the energy loss remains approximately constant throughout the depth of seawater. Between these two limits, there is a transition region in which the variation of the energy deposition per unit length should be taken into account. We describe our treatment of this regime in app~\ref{App:lines}.
In short, accounting for this variation shifts the detection efficiency curves to higher $dE/dz$ values, similar to increasing the pressure threshold (see fig.~\ref{fig:detection_efficiency_dEdzvary}).

Figure~\ref{fig:acoustic_contraints} presents the projected sensitivity in the spin-independent cross-section--dark matter mass plane. The red filled region shows the parameter space accessible to an array with a $45 \times 45 \times 10$ grid of hydrophones throughout $100\,\mathrm{km^3}$, assuming a $5\,\mathrm{mPa}$ threshold and requiring signals in at least ten hydrophones.
When the detection efficiency falls below 10\%, we set it to zero to ensure that only regions with sufficiently robust signal detection contribute to the sensitivity.

The right boundary of the red region in fig.~\ref{fig:acoustic_contraints} is set by the dark matter flux, while the lower boundary is determined by the detection threshold. For a $35\,\mathrm{mPa}$ threshold, the lower boundary would shift upward by approximately a factor of seven, consistent with our detection efficiency analysis (cf.\ fig.~\ref{fig:detection_efficiency}). The diagonal edge is set by our requirement that $\ell_{\rm{sea}}>20\,\mathrm{m}$, while the transition at $m_{\chi}\sim10^{-7}\,\mathrm{g}$ arises because the energy deposition begins to vary significantly with depth, requiring larger signal amplitudes for detection.

The acoustic array provides sensitivity to previously unexplored parameter space in the regime $m_{\chi} \sim 10^{-3}\,\mathrm{g}$ and $\sigma_{\chi}\sim10^{-8}\,\mathrm{cm}^2$,
and provides complementary coverage to existing constraints from cosmology and astrophysics (grey), non-traditional search strategies or repurposed experimental data (blue), and underground terrestrial experiments (green). 

We make two final comments. Firstly, a hydrophone array is sensitive to both spin-independent interactions through scattering with oxygen and hydrogen nuclei, and spin-dependent interactions through hydrogen, unlike many existing constraints that only probe spin-independent interactions. Secondly, for point-like candidates, the mean free path in seawater cannot be smaller than the intermolecular distance ($\sim 3 \times 10^{-10}$\,m), which leads to an energy deposition per unit length of $dE/dz \sim 10^{-4}\,\mathrm{keV}/\mathrm{fm}$. From fig.~\ref{fig:detection_efficiency}, this is below our detection threshold. However, as noted in ref.~\cite{Digman:2019wdm}, the large cross-sections probed in this analysis require dark matter to be composite, and composite candidates naturally produce larger energy depositions over extended regions, making them well-suited as targets for acoustic detection.

\section{Summary} \label{sec:conclusion}

The detection of ultra-heavy dark matter presents an experimental challenge because of its extremely low flux, which scales inversely with mass. For candidates heavier than approximately $10\, \upmu \mathrm{g}$ ($\sim 10^{19}\,\mathrm{GeV}$), the detector volumes required exceed those achievable with conventional underground direct detection experiments. Alternative detection strategies using geological, atmospheric, and astrophysical targets have been proposed to overcome this~challenge. 

In this work, we have explored the potential of another avenue: large underwater acoustic arrays to detect thermo-acoustic signals generated by ultra-heavy dark matter as it traverses seawater (cf.\ fig.~\ref{fig:idea}). These proposed arrays, which could instrument volumes of order $100\,\mathrm{km}^3$ with relatively sparse instrumentation, were originally proposed for the detection of ultra-high energy neutrinos.

We first derived the thermo-acoustic signal induced by ultra-heavy dark matter scattering off hydrogen and oxygen nuclei,
carefully characterising the energy deposition density along the dark matter's track through seawater. Solving the acoustic pressure wave equation, we showed that attenuation-free solutions are bipolar in shape, with a~$\mathcal{O}(\mathrm{ns})$ duration (cf.\ fig.~\ref{fig:PressureSourceTime}) and broadband frequency content with a cut-off at $\sim 10^{10}\,\mathrm{Hz}$ (cf.\ fig.~\ref{fig:PressureSourceFreq}).

We then investigated attenuation in seawater arising from both viscous absorption in pure water and chemical relaxation effects from dissolved salts. While the pressure solutions maintain their bipolar shape, a new frequency cut-off was introduced in the $10$~--~$100\,\mathrm{kHz}$ band (cf.\ fig.~\ref{fig:PressureAttenFreq}), dependent on the distance between the source of the acoustic pulse and the hydrophone. This led to pressure pulses with significantly reduced amplitude and increased duration of $\mathcal{O}(\mathrm{\upmu s})$ (cf.\ fig.~\ref{fig:PressureAttenTime}), while also exhibiting greater asymmetry between compression and rarefaction peaks (cf.\ fig.~\ref{fig:RoverC}).

Finally, after characterising the dark matter flux through the array and the detection efficiency in terms of the energy-deposition per unit length (cf.\ figs.~\ref{fig:FluxModulation} and~\ref{fig:detection_efficiency}), we determined the sensitivity reach for a large-scale hydrophone array with volume $100\,\mathrm{km}^3$, instrumented with approximately 200 hydrophones per km$^3$ and assuming detection thresholds of $5\,\mathrm{mPa}$ and $35\,\mathrm{mPa}$ -- parameters similar to those proposed for ultra-high energy neutrino detection. Our projections (cf.\ fig.~\ref{fig:acoustic_contraints}) show that such an array could probe previously unexplored regions of parameter space. A distinctive feature of the acoustic detection method is its sensitivity to both spin-independent interactions through oxygen nuclei and spin-dependent interactions through hydrogen nuclei, contrasting with some existing searches that probe only spin-independent couplings.

To summarise, large underwater acoustic arrays, originally designed for ultra-high energy neutrino detection, could also serve as probes of ultra-heavy dark matter.
While the focus of this paper has been on theoretical signal characterisation and sensitivity projections, with a view to informing future detector designs, analysis of existing hydrophone data may already constrain parts of the parameter space. Such data includes that from SAUND~\cite{vandenbrouckeExperimentalStudyAcoustic2005} and SAUND~II~\cite{kurahashiSearchAcousticSignals2010}. We leave this for future work.

\begin{acknowledgments}
We thank Patrick Knights and Tim Marley for discussions on the use of SRIM. DC acknowledges support from a Science and Technology Facilities Council (STFC) Doctoral Training Grant. CAJO is supported by the Australian Research Council under the grant number DE220100225.
For the purpose of open access, the authors have applied a Creative Commons Attribution (CC BY) license to any Author Accepted Manuscript version arising from this submission. 
The data supporting the findings of this study are available within the paper. No experimental datasets were generated by this research. 
\end{acknowledgments}

\appendix
\section{Energy deposition density} \label{App:length}
In this appendix, we provide details of the steps leading to eq.~\eqref{eq:q_definition}.
As the dark matter passes through material~`$A$', it continuously loses energy through elastic scattering with nuclei according to $\delta E_{\chi}=- \overline{E}_A \delta z / \lambda_A$, 
where $\overline{E}_A$ is the mean nuclear recoil energy and $\lambda_A = 1/(n_A \sigma_{\chi})$ is the mean free path, expressed in terms of $n_A$ and $\sigma_{\chi}$, the number density and cross-section (assumed independent of the target species).

When traversing through seawater, the total energy loss is therefore
\begin{equation}\label{eq:dEdzgeneral}
\frac{d E_{\chi}}{d z} = - \sum_A \overline{E}_A n_{A} \sigma_{\chi}\;,
\end{equation}
where~$A=\{\mathrm{H},\,\mathrm{O} \}$ labels the atomic species, either hydrogen or oxygen, and the total mean free path is $\lambda_{\chi}^{-1} = \sum_A \lambda_A^{-1}=3 n_{\rm{sea}} \sigma_{\chi}$, where $n_{\rm{sea}} = n_{\rm{O}}=n_{\rm{H}}/2$.

For isotropic scattering, the differential cross-section per unit energy deposited into species $A$ can be parametrised as
\begin{equation}\label{eq:dXsecdEr}
\frac{d \sigma_A}{d E_A} = \frac{\sigma_{\chi}}{E_{A|\text{max}}}\;,
\end{equation}
where $E_{A|\text{max}}= 4 \mu^2_{A} E_{\chi}(z)/(m_A m_{\chi})$ and~$m_A$ and~$\mu_A$ are the target mass and the target-dark matter reduced mass, respectively, and $E_{\chi}(z)$ is the dark matter kinetic energy at position~$z$ in the medium. From this, we determine that the mean nuclear recoil energy for species~$A$ is $\overline{E}_A = 2 \mu_A^2 E_{\chi}(z)/(m_A m_{\chi})$. Solving eq.~\eqref{eq:dEdzgeneral} with this value of $\overline{E}_A$, we arrive at eq.~\eqref{eq:dEdz} when we assume: $\mu_A \approx m_A$ since $m_{\chi}\gg m_A$; the mass of a water molecule is $m_{\rm{sea}}\approx m_{\rm{O}}+2 m_{\rm{H}}$, where $m_{\rm{O}}$ and $m_{\rm{H}}$ are the oxygen and hydrogen atomic masses, respectively;
and the kinetic energy of the dark matter when it enters the water is $E_{\chi}(z=0)= m_{\chi}v_{\chi}^2/2$.

At this point, we should pause and consider the potential impact of form factors, which are absent from eq.~\eqref{eq:dXsecdEr}.
There are two form factors that could be included: the nuclear form factor and a form factor associated with the composite nature of the dark matter.

For spin-independent scattering, the nuclear form factor is well modelled by the Helm form factor~\cite{Helm:1956zz, Lewin:1995rx}. We only consider scattering from light nuclei, so
the impact when the Helm form factor is included is essentially negligible, since it leads to a less than~1\% change in the mean nuclear recoil energy.
Therefore, our results can be compared directly with the `Model 1' scenario defined in~ref.~\cite{adhikariFirstDirectDetection2022}, or with the wider set of constraints in refs.~\cite{Dvorkin:2013cea, grahamWhiteDwarfsDark2018, sidhumacroscopicDarkMatter2019, sidhuDeathSeriousInjury2020, bhoonahDetectingCompositeDark2021,Acevedo:2021tbl, cappielloNewExperimentalConstraints2021, dhakalNewConstraintsmacroscopic2023}, which follow the same parametrisation as that employed here.

For composite dark matter, we would generally expect an additional model-dependent form factor, which could, depending on the dark matter model under consideration, lead to a change in the resulting phenomenology.
We will not consider this further in this paper, but refer to the literature~\cite{Gelmini:2002ez, Laha:2013gva, Hardy:2015boa, Butcher:2016hic, grabowskaDetectingDarkBlobs2018, coskunerDirectDetectionBound2019} for a discussion of the possibilities that have been explored elsewhere.

\begin{figure}[t!]
    \centering
    \includegraphics[width=0.9\columnwidth]{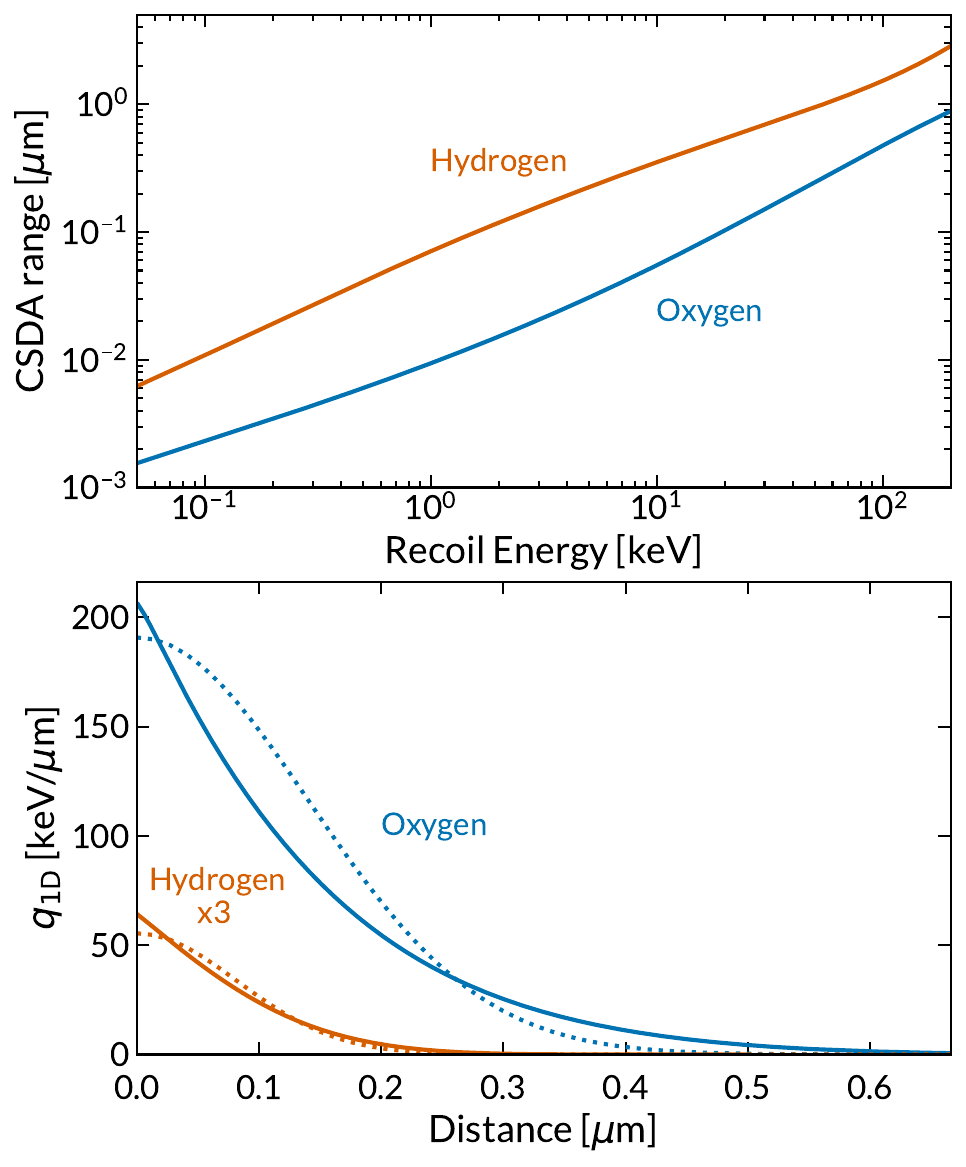}
    \caption{\label{fig:dEdx} Upper panel: The continuous slowing-down approximation range of hydrogen (red) and oxygen (blue) ions as a function of their recoil energy, calculated using the SRIM software for seawater (modelled as water with a density of~$\rho_{\rm{sea}}$). Lower panel: The one-dimensional energy deposition profile for oxygen and hydrogen recoils. Solid lines show the numerical calculation using eq.~\eqref{eq:avedEdz}, while dashed lines show a Gaussian distribution approximation. }
   \end{figure}

To calculate the energy deposition density, we need to relate the energy loss of the dark matter to the energy deposition in seawater.
We use the SRIM software package~\cite{Ziegler:2010bzy} to find the linear stopping power, $S(E) = -dE/dx$, of an oxygen and hydrogen ion as a function of recoil energy in water with the density set to~$\rho_{\rm{sea}}$.
By integrating the reciprocal stopping power over energy, we find the continuous slowing down approximation (CSDA) range.
The upper panel of fig.~\ref{fig:dEdx} shows this distance for hydrogen and oxygen ions as a function of the initial recoil energy.

With the standard scattering formula (e.g.,~ref.~\cite{Baudis:2012ig}),
\begin{equation}
\frac{dR_A}{dE_A} = \frac{\rho_{\chi}}{ m_{\chi} m_A} \int_{v_{\rm{min}}} d^3v_{\chi} \,f_{\chi}(v_{\chi}) \, v_{\chi} \,\frac{d\sigma_A}{d E_A}\;,
\end{equation}
where we follow ref.~\cite{Baxter:2021pqo} to model the dark matter velocity distribution $f_{\chi}(v_{\chi})$ (calculated with results from refs.~\cite{McCabe:2013kea,Evans:2018bqy}),
we find that the mean recoil energy of a hydrogen and oxygen ion, after averaging over the  dark matter velocity, is $\langle E \rangle_{\rm{H}}=1.9\,\mathrm{keV}$ and $\langle E\rangle_{\rm{O}}=30.2\,\mathrm{keV}$, respectively. From the upper panel of fig.~\ref{fig:dEdx}, we immediately learn that the typical distance scale for the recoils is~$\mathcal{O}(0.1\,\upmu\mathrm{m})$. This also implies that the energy deposition due to a single recoil can be treated as instantaneous with respect to hydrodynamic time scales, which are on the order of tens of milliseconds~\cite{Bevan:2009kd}.

We obtain the expected 1D energy deposition density, $q_{\rm{1D}}(z)$, as a function of distance by calculating
\begin{equation} \label{eq:avedEdz}
q_{\rm{1D}}(z) = \int dE' \frac{1}{R_A} \frac{dR_A}{dE'} \frac{dE}{dz} \left(E',z\right)\;.
\end{equation}
The results are shown as solid lines in the lower panel of Fig.~\ref{fig:dEdx} for hydrogen and oxygen ions.
We approximate these distributions with a Gaussian distribution (dashed lines), since this allows for an analytical solution to eq.~\eqref{pressure_equation_full}. When fitting the Gaussian function, we ensure the distributions integrate to match the mean energy obtained from eq.~\eqref{eq:avedEdz}.
For hydrogen, we find a standard deviation of~$\sigma_{\rm{recoil,H}}=0.082~\upmu\mathrm{m}$, while for oxygen,~$\sigma_{\rm{recoil,O}}=0.14~\upmu\mathrm{m}$.

Although the Gaussian fit slightly underestimates the numerical calculation from eq.~\eqref{eq:avedEdz} at both small and large distances,
this level of accuracy is sufficient for our purposes. This is because, as discussed in sec.~\ref{sec:Attenuated}, the high-frequency components of the signal --  those dependent on the precise shape of the energy deposition density -- are naturally attenuated in seawater.
Therefore, the Gaussian fit captures the essential characteristics of the energy deposition profile required for the current analysis. If higher precision were needed, the energy deposition density could be modelled as a linear combination of Gaussian functions, following the approach developed in quantum chemistry~\cite{Whitten1963}. This approach is effective since the solutions to the pressure wave equation, eq.~\eqref{eq:generalwave}, satisfy the principle of superposition.

To determine the 3D distribution, we apply the assumption of isotropic scattering. For a single collision, this translates to a product of Gaussian distributions in the $x$, $y$ and $z$ directions, normalised to give the mean energy per collision. 
For dark matter that traverses the medium along the $z$ direction, scattering occurs on average every mean free path $\lambda$, resulting in a total energy deposition represented by a sum of Gaussian distributions separated by $\lambda$. 
Assuming an infinitely long track and accounting for both hydrogen and oxygen contributions separately, we obtain
\begin{equation}
\begin{split}\label{eq:qr_full}
q(\bm{r}) &= \sum_A \langle E \rangle_A\, \frac{1}{2 \pi \sigma_{\mathrm{recoil},A}^2} \exp{\left( -\frac{x^2+y^2}{2 \sigma_{\mathrm{recoil},A}^2} \right)} \\
& \times \frac{1}{(2 \pi \sigma_{\mathrm{recoil},A}^2)^{1/2}} \sum_{k=-\infty}^{\infty} \exp{\left(-\frac{(z-k \lambda_A)^2}{2 \sigma_{\mathrm{recoil},A}^2}\right)}.\\
\end{split}
\end{equation}
Since $\lambda_A \ll \sigma_{\mathrm{recoil},A}$, we can approximate the sum over $k$ as an integral over $z'$, where $z' = k \lambda_A$, and by identifying $dE_A/dz= \langle E  \rangle_A / \lambda_A$, as used at the starting point of this appendix, we find 
\begin{equation}\label{eq:q_approx}
q(\bm{r}) \approx \sum_A \frac{1}{2 \pi }\frac{dE_A}{dz} \frac{1}{\sigma_{\mathrm{recoil},A}^2} \exp{\left(-\frac{\rho^2}{2 \sigma_{\mathrm{recoil},A}^2}\right)}\;,
\end{equation}
where $\rho = x^2+ y^2$. This matches the form given in eq.~\eqref{eq:q_definition} when $\sigma_A=\sigma_{\mathrm{recoil},A}$. Finally, if the dark matter radius~$R_{\chi}$ is larger than~$\sigma_{\mathrm{recoil,O}}$ and~$\sigma_{\mathrm{recoil,H}}$, the Gaussian width parameter in the~$x,y$ directions should be replaced with~$R_{\chi}$ in eq.~\eqref{eq:qr_full}, leading to the same simplified expression in eq.~\eqref{eq:q_approx} but with $\sigma_A = R_{\chi}$.

\section{Finite-width acoustic pressure solutions} \label{sec:pressure_derivations}

Our aim is to find a solution $p(\bf{r},t)$ of eq.~\eqref{pressure_equation_full} given the geometry of the energy deposition density from ultra-heavy dark matter, namely, as a long straight cylinder that is symmetric around the azimuthal angle. The energy density is given by $q(\bm{r'},t')=q(\bm{r'}) \Theta \left(t'\right)$, where $q(\bm{r'})$ is defined in eq.~\eqref{eq:q_definition}, and $t'=t-|\bm{r} - \bm{r'}|/c_s$.

It is useful to work in cylindrical coordinates where $z'$ is aligned along the centre of the energy deposition, and we align our coordinate system so that $\bm{r}=(x,y=0,z)$. We define a time $t_0$, according to $c^2_s t^2_0 = (x-x')^2+(y-y')^2 = (\rho-\rho' \cos\phi')^2+(\rho' \sin\phi')^2$, such that the propagation time $\tau$ satisfies $c^2_s \tau^2 =|\bm{r} - \bm{r'}|^2 = (z-z')^2+c_s^2 t_0^2$. 

This allows us to trade the $z'$ coordinate for $\tau$. This is useful since, when we act with one of the time derivatives from eq.~\eqref{pressure_equation_full} on $\Theta \left(t'\right)$ we get the term $\delta(t-\tau)$, so can immediately integrate over the $\tau$ (equivalent to $z'$) extent of the track. Upon doing this, we arrive at
\begin{widetext}
\begin{equation}
\begin{split}
p(\bm{r},t) = \frac{\alpha}{4\pi c_p} \frac{\partial}{\partial t} \int_0^{\infty} \rho' d\rho' \int_0^{2\pi} d\phi' &
\left(
 \frac{1}{\sqrt{t^2 - t_0^2}} \,  q\left(\rho', \phi', z+c_s\sqrt{t^2-t_0^2}\right) \left[\Theta\left( t-t_0\right) - \Theta\left( t-\tau_b\right)  \right] \right. \\
&\quad +\left.  \frac{1}{\sqrt{t^2 - t_0^2}}  \, q\left(\rho', \phi', z-c_s\sqrt{t^2-t_0^2}\right) \left[\Theta\left( t-t_0\right) - \Theta\left( t-\tau_a\right)  \right]
 \right)\;,
 \end{split} \label{eq:generalresult}
\end{equation}
where the two terms arise from contributions from $z'<z$ and $z'\geq z$, and $\tau_a$ and $\tau_b$ are the arrival times for each end of the track.

Since observations occur far from the track axis, we can use the approximation
\begin{equation}
c_s t_0 \approx \rho - \rho' \cos \phi'\;,
\end{equation}
and make a change of variables from $\rho', \phi' \rightarrow \rho', t_0$. 
After carefully considering the domain of integration in the $(\rho', t_0)$ coordinate system, we arrive at the expression
\begin{equation}\label{eq:general_time}
\begin{split}
p(\bm{r},t) = \frac{ \alpha }{2\pi c_p} \frac{\partial}{\partial t} \int_0^{\infty}  \frac{c_s dt_0}{\sqrt{t^2 - t_0^2}} \int_{|\rho - c_s t_0|}^{\infty}  \frac{ \rho' \,d\rho' }{\sqrt{ \rho'^2-(\rho - c_s t_0)^2 }} &
\left(
 \,  q\left(\rho', t_0, z+c_s\sqrt{t^2-t_0^2}\right) \left[\Theta\left( t-t_0\right) - \Theta\left( t-\tau_b\right)  \right] \right. \\
&\quad +\left.  \, q\left(\rho', t_0, z-c_s\sqrt{t^2-t_0^2}\right) \left[\Theta\left( t-t_0\right) - \Theta\left( t-\tau_a\right)  \right]
 \right)\;.
\end{split}
\end{equation}
\end{widetext}

\subsection{Constant-energy-deposition time-domain solution}

For long tracks, we can take the limit $\tau_a,\tau_b \to \infty$ and further assume we are in the regime where $dE/dz'$ is independent of $z'$. Were this assumption not valid and $dE/dz'$ changed rapidly with $z'$, the dark matter particle would soon come to rest, so the assumption of a long track would not be valid. 

\begin{figure}[t]
    \centering
    \includegraphics[width=0.99\linewidth]{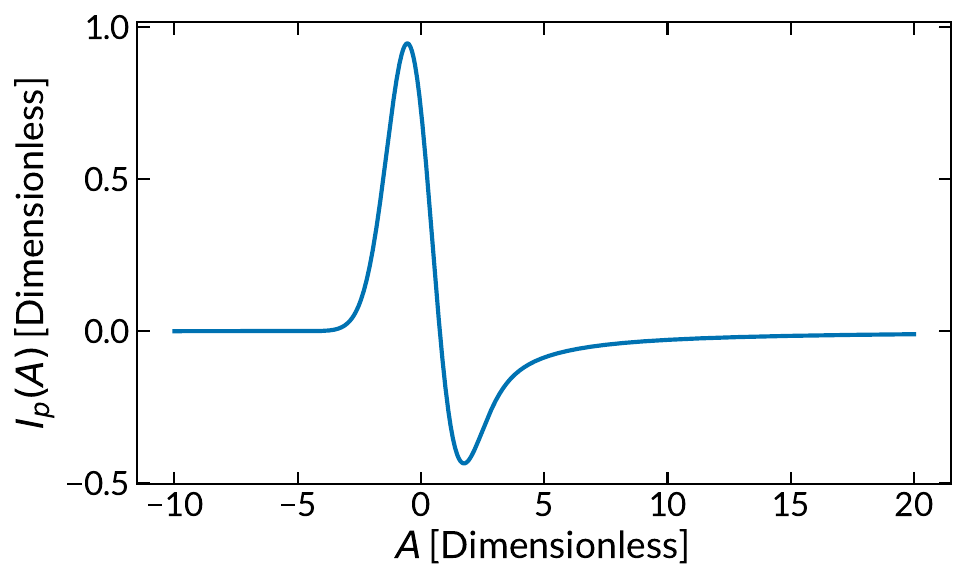}
    \caption{The dimensionless bipolar-pulse function $I_p(A)$ as a function of the dimensionless time parameter $A$. The function takes $\mathcal{O}(1)$ values, therefore the amplitude of the signal is determined by the prefactors.} 
    \label{fig:IA}
\end{figure}

After making these approximations and substituting eq.~\eqref{eq:q_definition} into eq.~\eqref{eq:general_time} (assuming one species contributes, so we can drop the subscript $A$), the integral over $\rho'$ can be performed analytically. However, the remaining integral has no (exact) closed-form solution, but it can be written in a form suitable for numerical integration by introducing $t_0 = t \sin u$, so that:
\begin{widetext}
\begin{equation}
p(\bm{r},t) = \frac{ \alpha}{2\pi c_p }  \frac{dE}{dz} \frac{1}{\sqrt{2 \pi }} \frac{c^2_s}{\sigma^3} \int_0^{\pi/2}  du \,\left(\rho-c_s t \sin u \right)\,\sin u\, \exp\left(-\frac{(\rho - c_s t \sin u)^2}{2 \sigma^2}\right)\;.
\label{eq:general_gaussian_solution}
\end{equation}

However, this form poses a significant computational challenge due to the extreme scales involved.
Specifically, recall that $\sigma\sim 0.1\, \upmu\mathrm{m}$ while $\rho\sim100\,\mathrm{m}$.
To explore alternative forms of the solution, we first use a trick of rewriting the exponential as an integral
\begin{equation}
\exp\left(-\frac{(\rho - c_s t \sin u)^2}{2 \sigma^2}\right) = \frac{1}{2 \pi \sigma^2} \int_{-\infty}^{\infty} dX \exp{\left( -\frac{\sigma^2 X^2}{2} \right)} \cos\left[ (\rho - c_s t \sin u) X \right]\;,
\end{equation}
which allows us to perform the integral over $u$, and to arrive at
\begin{align}\label{eq:general_gaussian_solution_alternative}
p(\bm{r},t) &= \frac{ \alpha}{2\pi c_p }  \frac{dE}{dz} \frac{c^2_s}{2}  \int_{0}^{\infty}dX\, \exp\left( -\frac{\sigma^2 X^2}{2}\right) X\left[\sin(X\rho)H_{-1}(c_s t X) - \cos(X\rho) J_{1}(c_s t X) \right]\;.
\end{align}
\end{widetext}
Here, $J_1(x)$ is the Bessel function of the first kind and $H_{-1}(x)$ is the Struve function, and we have used that the integrand is an even function. This is still an exact form, but as the integral is dominated by values of $X\sim\sigma^{-1}$, and since $\sigma \ll \rho$, the arguments of the Bessel and Struve functions are always large, so it is a good approximation to replace them with the leading term in their respective asymptotic expansions.
With this approximation, we arrive at the expression
\begin{widetext}
\begin{equation}\label{eq:p_asymptotic}
p(\bm{r},t; \sigma\ll\rho) = \frac{ \alpha}{2\pi c_p }  \frac{dE}{dz} \frac{c_s^2}{\sqrt{2 \pi \sigma^3}} \frac{1}{\sqrt{\rho}} \,I_p\left(\frac{t-\rho/c_s}{\sigma/c_s}\right)\;,
\end{equation}
where $I_p(A)$ is the `bipolar-pulse function', defined as
\begin{align} \label{eq:IA}
I_p(A) &= \int_{0}^{\infty}dY\,  \sqrt{Y} \exp\left( -\frac{Y^2}{2}\right) \,  \cos\left(A\, Y + \frac{\pi}{4} \right) \\
\begin{split}
 &=-\frac{\pi A}{4 \sqrt{2} (A^2)^{1/4}} \exp\left( -\frac{A^2}{4}\right) \left[ \left( A+\sqrt{A^2}\right)\left(I_{1/4}\left(\frac{A^2}{4}\right) - I_{3/4}\left(\frac{A^2}{4}\right) \right) \right.\\
 & \qquad \qquad  \qquad \qquad \qquad \qquad \qquad \qquad+ \left. \frac{\sqrt{2}}{\pi} \left( \sqrt{A^2} K_{1/4}\left(\frac{A^2}{4}\right) -A K_{3/4}\left(\frac{A^2}{4}\right)\right) \right]\;.
 \end{split}
\end{align}
\end{widetext}
Here, we have introduced $A=(c_s t-\rho)/\sigma$ and $Y=\sigma X$ as dimensionless variables, while
$I_{\nu}(z)$ and $K_{\nu}(z)$ are modified Bessel functions of the first and second kind, respectively. The variable $A$ can be considered as a dimensionless time parameter, while $Y$ can be regarded as a dimensionless wavevector. The variation of $I_p(A)$ as function of $A$ is shown in fig.~\ref{fig:IA}.
Upon taking the limits as $A \rightarrow \pm \infty$, we find the limiting forms given in eqs.~\eqref{eq:pulse_negative_t} and \eqref{eq:pulse_positive_t}. 

The asymptotic expansion provides an excellent approximation across all physically relevant distances. At $\rho=1\,\mathrm{m}$, eqs.~\eqref{eq:general_gaussian_solution} and~\eqref{eq:p_asymptotic} agree to within~$0.5\%$, while at larger distances the agreement improves dramatically: within better than~$5\times10^{-3}\%$ at $\rho=100\,\mathrm{m}$.

\subsection{Constant-energy-deposition frequency-domain solution}

The pressure pulse in the frequency domain is defined~as
\begin{equation}
\tilde{p}(\bm{r},\omega) = \int_{-\infty}^{\infty} dt \exp{\left( -i \omega t\right)} p(\bm{r},t)\;.
\end{equation}
Since $p(\bm{r},t)$ is a real-valued function, the negative frequency components are obtained from $\tilde{p}(\bm{r},-\omega)=\tilde{p}^{*}(\bm{r},\omega)$, i.e., from the complex conjugate of the positive frequency component. Therefore, we focus on the case when $\omega\geq0$.

To proceed, we find it most useful to use the form of $p(\bm{r},t)$ as given in eq.~\eqref{eq:general_gaussian_solution_alternative}.
By making use of the result that for $X,\omega>0$,
\begin{align}
&\int_{-\infty}^{\infty} dt\, e^{ - i \omega t} \left[\sin(X\rho)H_{-1}(c_s t X) - \cos(X \rho) J_{1}(c_s t  X) \right]  \\
&=\frac{2 i \omega}{c_s^2} \frac{1}{X \sqrt{X^2 - \omega^2/c_s^2}} \,e^{-i \rho X } \,\Theta\left(X- \frac{\omega}{c_s}\right) \;.
\end{align}
After letting $X=\omega/c_s \,\sec\theta$, we obtain the result given in eq.~\eqref{eq:p_freq_noatten}.


\section{Line-track acoustic pressure solutions} \label{App:lines}

In this appendix, we derive expressions for the acoustic pressure when the energy deposition occurs along a line track. For a line track, the energy deposition density takes the form
\begin{equation}\label{eq:align}
q(\rho', \phi', z')=\frac{1}{2 \pi} \frac{d E}{d z'}(z') \frac{\delta(\rho')}{\rho'}\;,
\end{equation}
where $\delta(\rho')$ is a Dirac delta function. As before, $(\rho, z)$ are the coordinates of the hydrophone, while the coordinates of the energy deposition are $(\rho', z')$.

We showed in sec.~\ref{sec:Acoustic_formalism} that in the observational frequency range, the acoustic pressure is independent of the track-width. Therefore, the line-track solution provides an appropriate limiting case for observational frequencies. 
We consider eq.~\eqref{eq:align} in two regimes: the constant-energy-deposition regime where $dE/dz'$ is assumed to be constant, allowing us to verify consistency with previous results, and the energy-deposition-varying regime, where we account for variation of $dE/dz'$ with position along the track.

\subsection{Constant-energy-deposition regime} \label{App:attenuation}
Upon substituting eq.~\eqref{eq:align} into eq.~\eqref{eq:generalresult} and taking the limits $\tau_a,\tau_b\to \infty$, corresponding to the infinite line case, we arrive at
\begin{equation}
\begin{split}
p(\rho,z,t) = \frac{\alpha}{4\pi c_p} \frac{\partial}{\partial t} &
\left\{
 \frac{\Theta\left( t-t_0\right)}{\sqrt{t^2 - t_0^2}} \left[ \frac{dE}{dz'}\left(z+c_s\sqrt{t^2-t_0^2}\right) \right. \right. \\
& \qquad \left. + \frac{dE}{dz'}\left(z-c_s\sqrt{t^2-t_0^2}\right) \right] 
\bigg\} \;.
\label{eq:longlinesolution}
\end{split}
\end{equation}

This form demonstrates that the pressure solution arises from the sum of two rays at distances $c_s \sqrt{t^2-t_0^2}$ along the track above and below the point $z$. For the regime where $dE/dz'$ remains constant, both contributions are equal, yielding
\begin{equation}
\begin{split}
    p(\rho, z , t) &= \frac{\alpha}{2\pi c_p} \frac{dE}{dz'} \frac{\partial}{\partial t}
    \left(\frac{\Theta\left( t-t_0\right)}{\sqrt{t^2 - t_0^2}} \right) \\
    &= \frac{\alpha}{2\pi c_p} \frac{dE}{dz'} \left(\frac{\delta(t - t_0)}{\sqrt{t^2 - t_0^2}} - \frac{t \Theta(t -t_0)}{(t^2 - t_0^2)^{3/2}} \right)\;.
\end{split}
\end{equation}
Since the energy deposition density has zero width, a Dirac delta function replaces the bipolar pulse shape at the initial arrival time $t_0=\rho/c_s$, followed by an amplitude that decays with a long tail. This result is consistent with that derived by Learned~\cite{learnedAcousticRadiationCharged1979}, differing only by a factor of two since the Learned calculation considers only rays below the observation point.

It is also consistent with the late-time expression given in eqs.~\eqref{eq:p_Afull} and~\eqref{eq:pulse_positive_t}. To demonstrate this, we write $t (t^2-t_0^2)^{-3/2}= t (t+t_0)^{-3/2} (t-t_0)^{-3/2}$. Since only small intervals around $t_0$ contribute significantly to the signal, so we can approximate $t (t+t_0)^{-3/2}\approx 2^{-3/2}(\rho/c_s)^{-0.5}$ to recover our earlier result.

\subsection{Energy-deposition-varying regime} \label{App:shorttrack}

\begin{figure}[t!]
    \centering
    \includegraphics[width=0.99\linewidth]{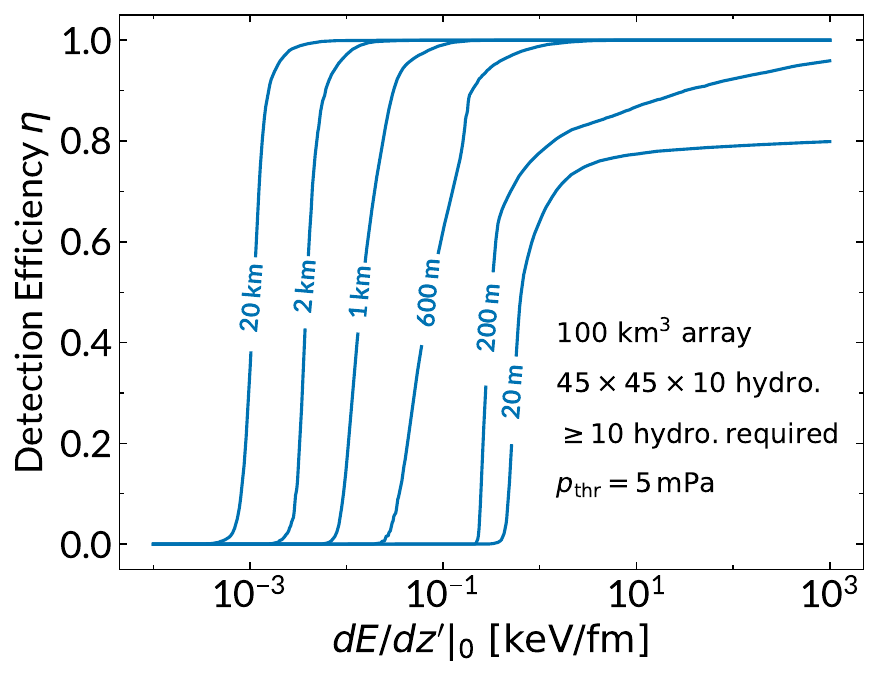}
    \caption{Detection efficiency as a function of the sea-surface energy deposition per unit length, $dE/dz'\big|_{0}$, shown for different characteristic energy-deposition length scales $\ell$. The detection efficiency is calculated for a $100\,\mathrm{km}^3$ array instrumented with $45 \times45 \times 10$ hydrophones, assuming a detection threshold of $5\,\mathrm{mPa}$ and requiring signals above threshold in at least ten hydrophones.
    }
    \label{fig:detection_efficiency_dEdzvary}
\end{figure}

We now consider the regime where $dE/dz'$ varies with position. For dark matter tracks, the energy deposition decreases exponentially along the track, motivating the form
\begin{equation}
\label{eq:qshorttrack}
q(\rho', \phi', z') = \frac{1}{2\pi} \frac{\delta(\rho')}{\rho'} \frac{dE}{dz'}\Big|_{0} \exp\left(-\frac{z'}{\ell}\right)\, ,
\end{equation}
where $dE/dz'|_{0}$ is the initial energy deposition rate and $\ell$ is the characteristic length scale over which the energy deposition varies. 

Substituting eq.~\eqref{eq:qshorttrack} into eq.~\eqref{eq:generalresult} and taking the limit $\tau_a,\tau_b\to \infty$, we obtain
\begin{equation}
\begin{split}
    p(\rho, z , t) &= \frac{\alpha}{2\pi c_p} \frac{dE}{dz'}\Big|_{0} \exp{\left(-\frac{z}{\ell}\right)} \\
    &\times \frac{\partial}{\partial t}
    \left[\frac{\Theta\left( t-t_0\right) }{\sqrt{t^2 - t_0^2}}\, \cosh\left(\frac{c_s \sqrt{t^2-t_0^2}}{\ell} \right)\right]\;.
\end{split} \label{eq:pshorttrack}
\end{equation}
The $\cosh$ term can be approximated as 
\begin{align}
&\cosh\left(\frac{c_s \sqrt{t^2-t_0^2}}{\ell} \right) \approx \cosh \left( \sqrt{\frac{t-t_0}{\Delta (\ell,\rho)}}\right)\;, \label{eq:cosh_approx} \\
&\Delta(\ell,\rho) = \frac{\ell^2}{2 \rho c_s} \approx 10^{4} \,\mathrm{\upmu s} \times\left(\frac{\ell}{100\,\mathrm{m}} \right)^2 \left( \frac{300\,\mathrm{m}}{\rho}\right)\;.
\label{eq:delta_def}
\end{align}
The timescale $\Delta(\ell,\rho)$ is much larger than the $\mathcal{O}(\mathrm{\upmu s})$ timescale over which the pressure pulse develops, so the $\cosh$ term varies slowly and can be approximated as unity. 
As a result, the exponential prefactor $\exp{\left(-z/\ell\right)}$, where $z$ is the distance along the track between the sea surface and the point closest to each hydrophone, provides the dominant modification in the regime of varying energy deposition.

In this regime, it is necessary to recalculate the efficiency to account for the exponential term.
Similarly to the procedure described in sec.~\ref{sec:projections}, we simulate $\mathcal{O}(10^5)$ tracks that intersect a plane of area $10.5 \,\mathrm{km} \times 10.5 \,\mathrm{km}$ through the array's centre. The top of the $45\times45\times10$ array is positioned $1.2\,\mathrm{km}$ below the sea surface. 
For each track and hydrophone in the array, we determine both $\rho$ and $z$, then calculate the acoustic pressure including the corresponding factor $\exp{\left(-z/\ell\right)}$. We assume that the sea surface is $1.2\,\mathrm{km}$ above the top of the array grid.

Figure~\ref{fig:detection_efficiency_dEdzvary} shows the detection efficiency as a function of the sea-surface energy deposition per unit length, $dE/dz'\big|_{0}$, for several values of $\ell$ between $20\,\mathrm{m}$ and $20\,\mathrm{km}$. In all cases, we assume that $p_{\mathrm{th}}=5\,\mathrm{mPa}$ and classify a track as detected if it produces a pressure that exceeds the threshold in at least ten hydrophones. Comparing with fig.~\ref{fig:detection_efficiency}, we see that we recover the constant-energy-deposition efficiency for $\ell\gtrsim20\,\mathrm{km}$. For smaller values of $\ell$, the attenuation of energy along the track means that we need larger values of $dE/dz'\big|_{0}$ to produce observable signals. However, we find that detection remains possible even at $\ell=20\,\mathrm{m}$, the smallest value we consider, although larger initial energy deposition rates are required.

\bibliographystyle{apsrev4-2}
\bibliography{acoustic_paper} 
\end{document}